\documentclass[sigconf]{acmart}
\usepackage[utf8]{inputenc}
\usepackage{enumitem}
\usepackage{rotating}
\usepackage{ccicons}          % For Creative Commons icons
\newcommand{\enquote}[1]{``#1''}

\copyrightyear{2026}
\acmYear{2026}
\setcopyright{cc}
\setcctype{by}
\acmConference[CHI '26]{Proceedings of the 2026 CHI Conference on Human Factors in Computing Systems}{April 13--17, 2026}{Barcelona, Spain}
\acmBooktitle{Proceedings of the 2026 CHI Conference on Human Factors in Computing Systems (CHI '26), April 13--17, 2026, Barcelona, Spain}
\acmDOI{10.1145/3772318.3791529}
\acmISBN{979-8-4007-2278-3/2026/04}

\AtBeginDocument{%
  }

\begin{document}

%% Load abstract and title from separate .tex
%% The "title" command has an optional parameter in square brackets,
\title{Reactive Writers: How Co-Writing with AI Changes How We Engage with Ideas}

\begin{abstract}
Emerging experimental evidence shows that writing with AI assistance can change both the views people express in writing and the opinions they hold afterwards. Yet, we lack substantive understanding of procedural and behavioral changes in co-writing with AI that underlie the observed opinion-shaping power of AI writing tools. We conducted a mixed-methods study, combining retrospective interviews with 19 participants about their AI co-writing experience with a quantitative analysis tracing engagement with ideas and opinions in 1{,}291 AI co-writing sessions. Our analysis shows that engaging with the AI's suggestions---reading them and deciding whether to accept them---becomes a central activity in the writing process, taking away from more traditional processes of ideation and language generation. As writers often do not complete their own ideation before engaging with suggestions, the suggested ideas and opinions seeded directions that writers then elaborated on. At the same time, writers did not notice the AI's influence and felt in full control of their writing, as they---in principle---could always edit the final text. We term this shift \textit{Reactive Writing}: an evaluation-first, suggestion-led writing practice that departs substantially from conventional composing in the presence of AI assistance and is highly vulnerable to AI-induced biases and opinion shifts.
\end{abstract}
\author{Advait Bhat}
\affiliation{%
  \institution{Paul G. Allen School of Computer Science, \\University of Washington}
  \city{Seattle}
  \state{Washington}
  \country{United States}
}
\email{advaitmb@cs.washington.edu}

\author{Marianne Aubin Le Quéré}
\affiliation{%
  \institution{Center for Information Technology Policy, \\Princeton University}
  \city{Princeton}
  \state{New Jersey}
  \country{United States}
}

\author{Mor Naaman}
\affiliation{%
  \institution{Cornell Tech}
  \city{New York}
  \state{New York}
  \country{United States}
}

\author{Maurice Jakesch}
\affiliation{%
  \institution{Bauhaus University}
  \city{Weimar}
  \country{Germany}
}
%\email{maurice.jakesch@uni-weimar.de}

\renewcommand{\shortauthors}{Bhat et al.}

%% CCSXML code




\maketitle

% Load content files
\section{Introduction}

AI writing assistants have rapidly become ubiquitous tools in everyday communication, from email composition to creative writing~\citep{hancock2020ai,Spataro2023Copilot,Grammarly2025About}. Systems like GPT-based autocomplete, Grammarly~\citep{Grammarly2025About}, and integrated assistants in Google Docs and Microsoft Word~\citep{Wu2018SmartCompose,Google2023DuetAI,Spataro2023Copilot} promise to augment human writing by increasing speed, improving surface-level quality, and democratizing access to polished prose~\citep{NoyZhang2023Science,BrynjolfssonLiRaymond2025}. These tools generate contextually relevant suggestions that can help writers overcome blank-page paralysis, correct grammatical errors, and produce professional-sounding text~\citep{yuan_wordcraft_2022, singh_where_2022}. As these systems become increasingly sophisticated and deeply embedded in our writing ecosystem, they are reshaping how millions of people express themselves in text.

However, emerging evidence reveals tradeoffs in this human-AI collaboration. Recent studies demonstrate that writers using AI assistance may experience cognitive offloading that hinders idea generation~\cite{Kosmyna2025YourBrainOnChatGPT, umarova2025problematic}, produce more predictable language~\cite{arnold_predictive_2020, buschek_impact_2021}, and report reduced satisfaction with the writing process despite improved productivity~\cite{dhillon2024shaping}. More troublingly, AI writing assistants have been shown to systematically influence not just how people write, but what they express---biased language models can shift both the opinions writers convey in their text and their subsequent personal attitudes~\cite{jakesch2023co, williams2025biased}, affect sentiment through positive bias in suggestion models~\cite{arnold_sentiment_2018, Hohenstein2020}, and nudge participants to write about themselves using different lenses~\cite{poddar2023ai}. These effects often occur without writers' awareness~\cite{jakesch2023co}, suggesting a form of influence that operates below conscious recognition.

While these outcome-level effects are concerning, understanding outcomes alone is insufficient: we need a descriptive account of how AI suggestions reshape the process and experience dimensions of writing. We lack detailed process-level accounts of how writers actually interact with opinionated suggestions during composition. How do writers encounter, evaluate, and integrate AI suggestions moment-by-moment? When suggestions appear inline during composition, how does this reshape the fundamental cognitive processes of ideation, memory retrieval, and expression~\cite{flower1981cognitive, Kintsch1980DiscourseProduction}? Without understanding these interactions and their cumulative effects on the writing process, we cannot fully grasp how AI assistance transforms human expression, or design systems that preserve writer agency.

The impact of such systems extends far beyond individual writing sessions. If AI suggestions systematically influence which ideas writers express, they may subtly direct public discourse toward particular topics while marginalizing others. As AI-assisted text proliferates across the web---from social media posts to professional documents---the biases and priorities embedded in language models become entrenched in our collective information ecosystem~\cite{bender2021dangers, johnson2022ghost, agarwal2025ai}. Understanding how AI shapes the writing process is therefore critical not just for individual writers, but also for addressing broader concerns about human expression and democratic discourse.
To address this gap, we investigate two interrelated research questions: 
\begin{enumerate}[label=\textbf{RQ\arabic*:}]
  \item How does an inline opinionated AI writing assistant reshape writers' processes of ideation and composition?
  \item How does co-writing with AI affect the ideas that writers ultimately express?
\end{enumerate}

We employ a mixed-methods approach. We conducted 19 cued retrospective verbal protocol interviews in which participants reflected on their writing process while viewing replays of their AI-assisted writing sessions. We complement these with a quantitative analysis of interaction logs from 1{,}291 writing sessions to examine systematic patterns in how AI suggestions propagate into final texts.

We examine these questions in the context of an opinionated, inline AI writing assistant. 
We focus on inline suggestions that appear directly at the cursor in tools like email clients and word processors, since this interface is the most common way that AI suggestions are incorporated into writing today~\citep{Wu2018SmartCompose,Google2023DuetAI,Spataro2023Copilot}.
Within this interface, we deploy an inline assistant powered by an opinionated language model that surfaces biased text completions, in this case taking a definite stance on whether social media is good or bad for society.
We deploy an opinionated model for three reasons. 
First, while most deployed autocomplete systems claim to avoid overt political content, even ostensibly neutral models encode systematic political and moral biases, and any assistant must still choose which framings, examples, or arguments to present~\citep{feng2023pretraining}.
The mechanisms by which writers encounter, evaluate, and integrate AI suggestions operate regardless of whether bias is explicit or implicit. 
Second, we treat an explicitly opinionated assistant as an analytic extreme, which allows us to more clearly observe these mechanisms.
This approach has been deployed successfully for studying confident versus uncertain AI~\citep{kadoma2024role} and culturally aligned versus misaligned AI~\citep{agarwal2025ai}.
Third, prior work suggests cognitive and persuasive impacts of using biased AI writing assistants~\cite{williams2025biased, jakesch2023co}, and we seek to understand the processes underlying these effects.

Our findings reveal a reorientation of the writing process that we term \textbf{reactive writing}: a mode of writing where writers shift from generating ideas through memory retrieval to evaluating and elaborating on AI-presented content. Through detailed process accounts, we document how rapid inline suggestions interrupt ideation, redirecting writers' attention toward agreement-based evaluation rather than independent idea generation. Writers maintain a strong sense of agency and control even as they accept suggestions that introduce new topics and framings into their text. Our quantitative analysis confirms that exposure to AI suggestions on specific topics strongly predicts their presence in final essays, with effects persisting even when controlling for directly accepted text. Together, these findings illuminate how AI writing assistants can function as subtle but powerful algorithmic agenda-setting technologies~\cite{trielli2022algorithmic, mccombs1972agenda}, shaping not just how we write, but fundamentally what we think and write about.
\section{Related Work}
We situate our work within research on technologies that support writing and emerging work on the persuasive effects of AI and algorithms.

\subsection{Writing with Technological Assistance}

Classical theories of writing posit that writing begins with human ideation, where \textit{``the generating process derives its first memory probe from information about the topic and the intended audience.''}~\cite{Kintsch1980DiscourseProduction, flower1981cognitive}. These models, such as \citet{flower1981cognitive}'s Cognitive Process Model of Writing, identify key internal writing processes including the proposer (generating ideas), translator (converting ideas to language), transcriber (producing written text), and evaluator (assessing outputs) as the building blocks of the writing process.

With the evolution of writing technologies, researchers have tried to augment the writing process using tools for improving writing quality or productivity. HCI research on text entry has traditionally focused on efficiency~\cite{kristensson2014inviscid}, with systems providing word- or sentence-level suggestions based on likelihood distributions~\cite{Vertanen2015velocitap, Bi2014, arnold2016suggesting, buschek_impact_2021}. These systems typically suggested one to three subsequent words~\cite{Dunlop2012, Fowler2015, Gordon2016, Quinn2016chi}, with more recent systems providing multiple short reply suggestions~\cite{Kannan2016smartreply} or single longer phrases~\cite{chen2019gmail}. Extensive suggestions were generally avoided because they required time to read and select, and studies showed that features such as autocorrection and word prediction could negatively impact writing performance~\cite{Banovic2019mobilehci, Dalvi2016, Buschek2018researchime, Palin2019}.

However, as computational capabilities have advanced, particularly with large language models~\cite{winata2021language, bommasani2021opportunities, vaswani2017attention}, there has been a shift toward viewing writing assistants as active partners rather than tools for writing faster~\cite{lee_coauthor_2022, daijin_yang_ai_nodate, yuan_wordcraft_2022}. Such systems have been integrated and studied as tools supporting idea generation~\cite{lee_coauthor_2022, singh_where_2022, yuan_wordcraft_2022, bhat2023Interacting}, story writing~\cite{singh_where_2022, yuan_wordcraft_2022}, text revision~\cite{Cui2020, Zhang2019}, and creative writing~\cite{clark2018creative, Gero2019MetaphoriaAA}. Recent work has mapped design spaces for intelligent writing assistants, identifying five pertinent aspects: the user, task, interaction, technology, and ecosystem~\cite{lee2024design}. \citet{reza2025co} distinguish different writing assistants based on their design approach, including tools that provide structured guidance, critical feedback, exploration, and active co-writing. AI autocomplete tools, such as the one used in this study, fall under the category of active co-writing assistants.

Research has also examined how users actually engage with AI writing assistance and how it may affect the process of writing. \citet{bhat2023Interacting} discuss how writers evaluate suggestions and integrate them into cognitive writing processes such as idea generation, language generation, and transcription. \citet{singh_where_2022} observe writers making ``integrative leaps'' when incorporating AI suggestions during creative story writing with multimodal systems. \citet{buschek_impact_2021} identify nine behavioral patterns of interaction with suggestions, ranging from complete avoidance to chaining multiple suggestions together. This engagement can take different forms: writers can leverage AI assistants in ways that foster ideation and creativity, and potentially engender constructive learning, or in ways that focus on rote and mindless tasks~\cite{umarova2025problematic}. Writers' perceptions of such tools vary, with different writers being comfortable delegating or maintaining ownership over different writing processes such as planning and ideation~\cite{reza2025co}.

Beyond affecting the writing process, AI assistance has also been shown to impact the products of writing. AI suggestions have been shown to produce more predictable language and shorter sentences~\cite{arnold_predictive_2020}, increase the use of standard phrases~\cite{buschek_impact_2021, bhat2023Interacting}, and affect sentiment, with positive sentiment bias in suggestion models leading to more positive writing~\cite{arnold_sentiment_2018, Hohenstein2020}. Additionally, AI assistants can nudge participants to write about themselves using different lenses~\cite{poddar2023ai}.

Empirical work has revealed important trade-offs in AI-assisted writing. Increased scaffolding with active co-writing tools such as autocomplete AI suggestions can indeed improve writing quality and productivity; however, this may come at the cost of writers' satisfaction with the process~\cite{dhillon2024shaping}. Recent work has suggested that making AI suggestions reflect participants' own voice can improve feelings of control and agency~\cite{kadoma2024role}.

\subsection{Influence, Algorithms, and AI Persuasion}
Social influence alters individuals' thoughts, feelings, and actions through interactions with others~\cite{rashotte2007social}.
The impacts of such influence range from collaboration to unethical behavior at the individual level~\cite{asch1951effects, milgram1963behavioral}, and from effects on markets, voting, and health behaviors at the societal level~\cite{shiller2015irrational, lazarsfeld1968people, christakis2007spread, christakis2008collective}. 
Sources of influence vary widely, including friends and family, experts, and internet celebrities~\cite{goel2012structure, marwick2011tweet}. 

In digital environments, influence increasingly comes from algorithmic systems. 
Online, influence can come from brand pages and bots~\cite{ferrara2016rise}, as well as technical artifacts such as recommender systems and chatbots that steer user decisions~\cite{berkovsky2012influencing, leonard2008richard, cosley2003seeing, gunaratne2018persuasive}. 
Algorithmic influence depends on user perceptions and trust~\cite{logg2019algorithm, gunaratne2018persuasive}, and people may rely more on algorithms in uncertain environments~\cite{bogert2021humans}. 
Growing public awareness of AI has led many to view algorithms as authoritative, which can foster over-reliance even when algorithms are wrong~\cite{10.1145/3491102.3517533, logg2019algorithm, araujo2020ai, parasuraman1997humans, parasuraman2010complacency, wickens2015complacency}.

In the context of writing assistance, emerging work examines how algorithmic systems may influence writers through suggested content. 
Text suggestions based on language models can inherit biases and dominant views from training corpora~\cite{basta2019evaluating, kurita2019measuring, sheng2019woman, hutchinson2020social}, and the degree of alignment between writers' perspectives and model biases may affect interaction patterns. 
These writing assistants represent a specific instance of algorithmic influence, where the act of providing suggestions can shape writing content and direction~\cite{arnold_sentiment_2018}.
More broadly, the political implications of social media and recommender systems have received extensive scrutiny~\cite{zhuravskaya2020political}. 
Technologies that shape public opinion are vulnerable to powerful political and commercial interests~\cite{bradshaw2017troops}. 
As algorithms become constitutive features of public life, they can shift the political landscape~\cite{gillespie2014relevance, aral2019protecting} and contribute to polarization, even without being explicitly designed to change opinions~\cite{bruns2019filter, cinelli2021echo, bail2018exposure}.

Today, with the advent of large language models such as GPT-style systems~\cite{brown2020language, radford2019language}, emerging work has found that humans may be particularly susceptible to AI persuasion.
These models can produce human-like text~\cite{jakesch2022human}, giving rise to concerns over broad ethical and social risks~\cite{weidinger2021ethical, weidinger2022taxonomy}. 
New studies have found that chatbot-based interactions can perform remarkably well at human persuasion tasks, including through personalized arguments on socially contentious issues~\cite{costello2024durably, salvi2025conversational}.
In the context of AI writing assistants, previous work has also documented that opinionated models can consistently change people's minds across a range of social issues~\cite{williams2025biased, jakesch2023co}.
Prior work has shown that writing produced with biased AI suggestions reflects the AI's positive or negative opinions~\cite{jakesch2023co} and framing~\cite{poddar2023ai}, but we lack a granular understanding of how this process unfolds through the introduction of new ideas.

The products of human-AI co-writing have significant downstream impacts on our information ecosystem.
LLMs frequently mirror values in their training data, often aligning more with dominant U.S. values than those of other cultures~\cite{johnson2022ghost, agarwal2025ai}.
Since future AI models are likely to be trained on these human-AI co-written texts, biases may become entrenched over time, causing value lock-in or even model collapse~\cite{bender2021dangers, shumailov2024ai}.
Researchers have raised concerns that entrenching biases into predictive text models may produce effects where people feel differentially included~\cite{kadoma2024role}.

The convergence of writing assistance technologies with algorithmic influence and persuasion creates new concerns about how AI shapes both the cognitive processes of writing and the ideas writers ultimately express. 
Recent research reveals that interactions with AI writing assistants can lead writers to adopt cognitive shortcuts that undermine the learning benefits traditionally associated with writing. \citet{umarova2025problematic} demonstrate that problematic writer-AI interactions can hinder idea generation. Similarly,~\citet{Kosmyna2025YourBrainOnChatGPT} provide neurological evidence that LLM assistance creates ``cognitive debt,'' with participants showing weaker neural connectivity patterns and reduced cognitive engagement compared to those writing without AI support. 

These findings take on additional significance in light of~\citet{jakesch2023co}'s demonstration that opinionated language models can shift both what users write and their subsequent attitudes, often without users' awareness of the model's influence. Together, these studies suggest that as writers increasingly rely on AI assistance, they may simultaneously become more cognitively passive and more susceptible to the biases embedded in AI suggestions. However, existing research has largely examined these phenomena separately—studying either the cognitive impacts of AI assistance or its persuasive effects, but not how these processes interact during the act of writing itself.
\section{Methods}
\label{sec:methods}
The current study investigates two interrelated questions: how an inline opinionated AI writing assistant reshapes the writer's \textbf{process} of ideation and composition (RQ1), and how the AI shifts the \textbf{content} that writers ultimately produce (RQ2).
Since the first question concerns people's lived experiences, while the second relates to data in the text output of a large-scale experiment, we chose a mixed-methods design to answer these questions~\cite{creswell2017designing}.
In this methods section, we first review the experimental platform (Section \ref{subsec:writing_platform}), then introduce our qualitative (Section \ref{subsec:qual}) and quantitative methodologies (Section \ref{subsec:quant}).

\subsection{Experimental Platform}
\label{subsec:writing_platform}
To observe participants in the process of co-writing with AI, we needed an experimental platform that provided both fine-grained control over the writing interface and detailed logs of the interactions. 
We chose to work with the open-source AI-Cowriting Research Platform (AI-CRP) by~\citet{jakesch2024AICRP}. The platform combines a mock-up of a social media discussion page with a rich-text editor and an embedded inline opinionated AI writing assistant. To provide a realistic social media context, the platform replicates the design of a Reddit discussion page. 

After providing informed consent, participants were asked to compose an opinion statement on the platform expressing their perspective on the impact of social media on society. The discussion topic of social media was chosen because it is an accessible topic of personal and societal relevance, and would allow us to compare our results to data collected in previous studies \cite{jakesch2023co}. 
We connected the experimental platform to a GPT-3-based language model in the backend~\cite{brown2020language}, generating contextual suggestions that appeared after natural pauses in participants' typing (typically 2 seconds). 
To make the influence of the AI model on people's writing more salient so that the changes in participants' writing and ideation processes could be discerned more clearly, we prompted the model to produce opinionated suggestions. 
The platform captures comprehensive interaction logs including keystroke timing, cursor movements, suggestion appearances, and participant responses (accept, edit, or ignore).

\subsection{Qualitative Interviews}
\label{subsec:qual}

For the qualitative component, participants were recruited to use the AI-CRP tool~\cite{jakesch2024AICRP} and then complete a follow-up interview about their experience. 
We conducted cue-based retrospective interviews to understand writers' cognitive decision-making processes during composition with an opinionated AI writing assistant~\cite{ericsson1993protocolanalysis,gass2016stimulated}. We adopted a replay-based retrospective protocol inspired by Progression Analysis~\cite{GrésillonPerrin+2014+79+112, perrin2003progression} where we aligned verbal protocols with time-stamped traces of the session so that interpretations could be anchored in observable events rather than unaided recall.

\subsubsection{Participant Recruitment}

We recruited US- and UK-based participants aged 18 and above from Prolific~\cite{palan2018prolific}. Participants booked a single session via Calendly that included the writing task immediately followed by the interview; the interview occurred right after task completion within the same scheduled slot. We continued recruitment and data collection until we reached thematic saturation after 19 interviews—the point at which additional interviews yielded little to no new codes or themes relevant to our research questions \cite{guest2006howmany,hennink2017saturation}.

Participants were compensated at $\$30$ per hour for their time, covering both the writing task and interview components. 
Following each interview, participants received a full debriefing that disclosed the AI assistant's opinionated stance and the study's research objectives. 
All participants provided informed consent for both the writing task and interview recording before participation began. The study protocols were approved by the university's Institutional Review Board.

\subsubsection{Interview Protocol and Qualitative Analysis}

\begin{figure*}[ht]
    \centering
    \includegraphics[width=0.65\linewidth]{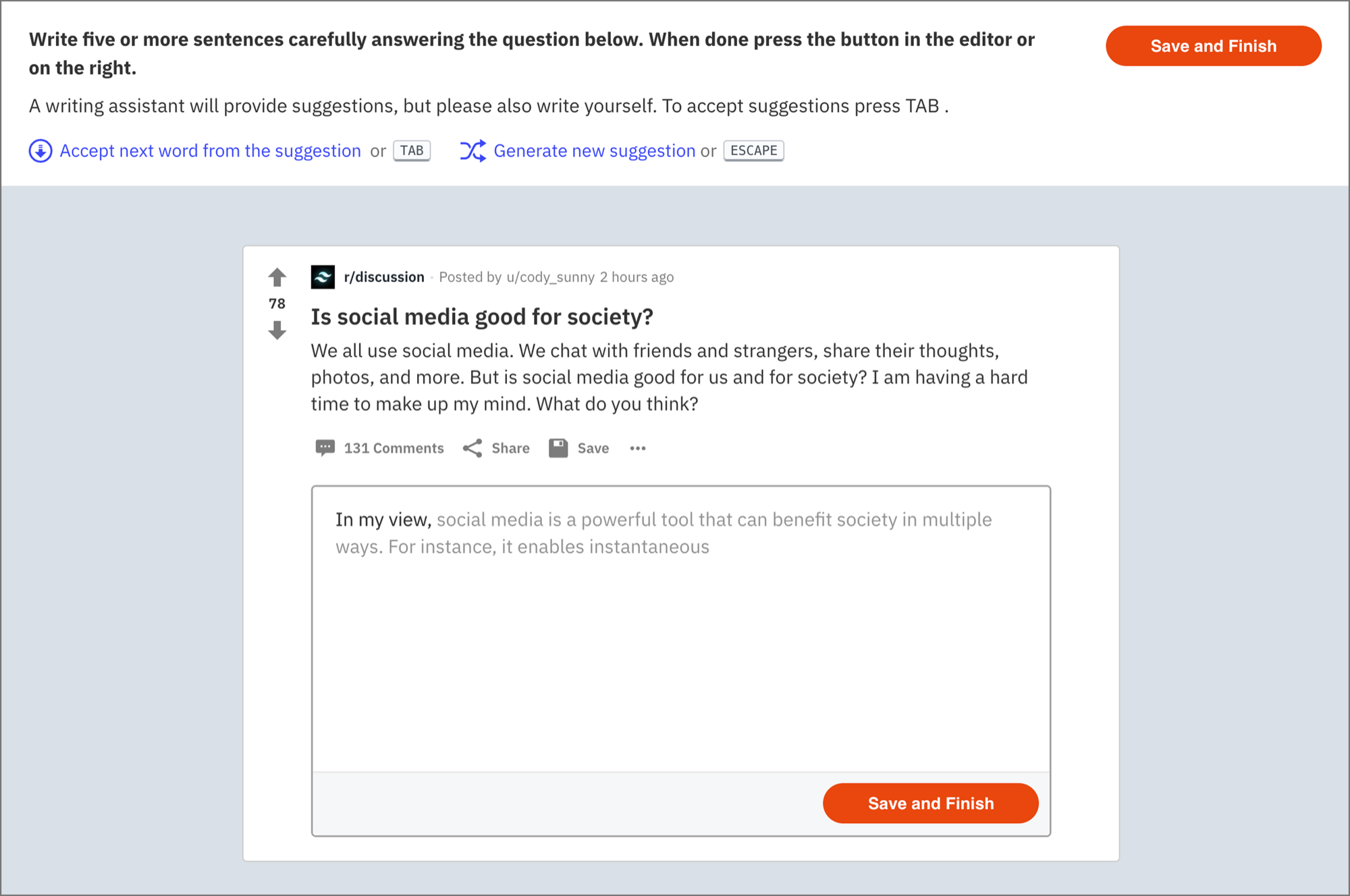}
    \caption{Screenshot of the experimental writing platform showing AI suggestions.}
    \label{fig:platform}
    \Description{The image is a screenshot of a Reddit-style discussion platform. At the top, a post asks, "Is social media good for society?" The user is composing a response in a text box, starting with "In my view" and the inline AI suggestions appear in gray saying "social media is a powerful tool that can benefit society in multiple ways. For instance, it enables instantaneous...". The interface includes buttons for saving and sharing and instructions on how to complete the task along with how to accept the AI suggestions.}
\end{figure*}

During their scheduled timeslot, participants were randomly assigned to write using an AI assistant with either a critical or positive stance towards social media, mirroring the treatment used in~\citet{jakesch2024AICRP} (see Figure~\ref{fig:platform}).
Participants received a link to the writing platform and completed the writing task in their assigned condition.
Immediately following task completion, participants joined a Zoom call for a 30--45 minute retrospective interview. 
This immediate transition allowed us to conduct the interview while the writing experience remained vivid in participants' memory, before any substantial memory decay could occur.

We developed a custom replay tool to use during the interview that reconstructed the participants' entire writing session from the interaction logs. 
The tool visualized the evolving user text, cursor position, and AI suggestions as they appeared in the text editor, with accepted suggestions clearly highlighted. The interviewer could play, pause, scrub through the timeline, and jump to specific moments in the session. Using this replay as a visual cue provided a concrete anchor for participants’ retrospective accounts, helping them ground recollections in specific, observable moments rather than relying solely on unaided memory \cite{lyle2003stimulated,calderhead1981stimulated,gass2016stimulated}. 
The semi-structured interview protocol was built around this replay of each participant’s writing experience with the AI tool. We elicited retrospective verbal protocols using a semi-structured interview guide \cite{kallio2016semistructured,dicicco-bloom2006qual} (Interview guide in appendix). We identified salient actions during the writing task (e.g., beginning to write, suggestions appearing, accepting/rejecting/editing suggestions). As the replay progressed, we paused at these actions and invited participants to narrate what they were thinking or doing in those specific moments. For example, when a suggestion appeared, we asked: ``What did this suggestion make you think?'' and ``What were you planning to write here?''

Interviews were transcribed using OtterAI and transcripts were manually refined for accuracy~\cite{otterai2025}. 
We coded the resulting verbal protocols against the interview video with the process replay to preserve the context of participants' actions~\cite{ericsson1993protocolanalysis, gass2016stimulated}.
We used thematic analysis inspired by grounded theory as a methodological starting point~\cite{braunandclarke}. We began with an initial open-coding approach~\cite{saldana2021coding, corbin2015basics} for the first 10 participants' protocols to develop a preliminary codebook. The open coding was conducted by one researcher.
We iteratively collated codes into candidate themes, reviewed them against the full corpus and the video recordings, and wrote analytic memos throughout to document emerging patterns and formulate preliminary theories~\cite{braunandclarke}.
After analyzing the initial 10 protocols, we conducted subsequent data collection and analysis simultaneously, allowing findings from earlier interviews to inform later data collection.

\subsection{Quantitative Topic Analysis}
\label{subsec:quant}

The quantitative analysis examines the downstream effects of AI autocomplete suggestions on written content at scale.
Prior work has established that opinionated writing assistants affect the opinions people write about and hold after writing \cite{jakesch2023co, williams2025biased}, but it is unclear how this influence unfolds. 
In the quantitative part of the study, we examine how the AI assistant influenced participants' engagement with ideas and concepts through a topic-based analysis.

\subsubsection{Dataset and Sample}
The analysis complements the data collected in interviews with an extensive dataset of interaction logs of 1{,}506 co-writing sessions publicly shared by~\citet{jakesch2023co}.
Collected from a comparable online experiment,
the interaction logs capture every user interaction with the text editor, including suggestions appearing, accepted suggestions, and text edits over time.
The dataset includes these detailed interaction logs, a unique participant identifier, and a label for the experimental condition. 

We removed malformed data, incomplete process captures, and outliers with unusually high or low writing session times and pause times.
After this process, 1{,}291 participants remained in our sample, with 462 in the no-AI control group, 423 in the pro-social media treatment group, and 406 in the anti-social media treatment group.

For an in-depth analysis of participants' interactions with the AI assistant, we needed an appropriately scoped unit of analysis.
To capture how AI suggestions influenced what participants wrote about, we needed to define our units of analysis carefully. For treatment groups, we labeled 'writing bursts'---text that a user had written between two pauses---along with the AI suggestions that appeared after each pause with topics. For the control group, we labeled sentences split from the essay column. These became our units of analysis---text segments that we could then analyze at the participant $\times$ topic level, aggregating across the full session for outcome and exposure measures.

\begin{figure}[htbp]
\centering\includegraphics[width=0.45\textwidth]{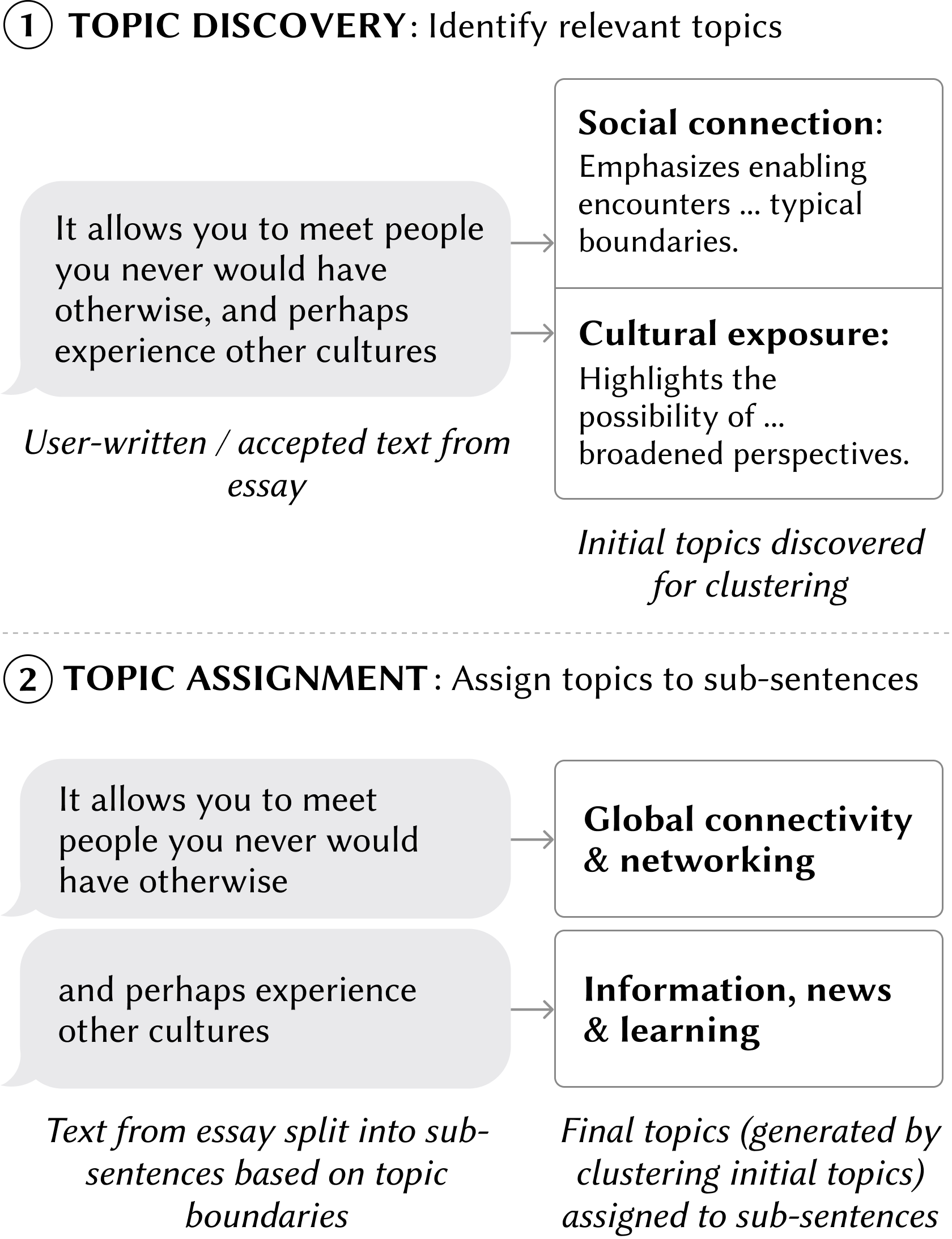}
    \caption{\textbf{Simplified overview of the topic classification pipeline.} Step 1 (\textit{topic discovery}) shows how initial topics were identified from user-written and accepted text using \texttt{gpt-3.5-turbo}, with each topic receiving a description to enable clustering. Step 2 (\textit{topic assignment}) shows how text was split into sub-sentences based on topic separation, with each sub-sentence assigned by \texttt{gpt-3.5-turbo} to one of the final topics derived from hierarchical clustering.}
    \Description{Two-step pipeline diagram using gpt-3.5-turbo. Step 1 (Topic Discovery): user-written text in a speech bubble leads to two boxes showing discovered topics with descriptions (Social connection, Cultural exposure). Step 2 (Topic Assignment): the same text split into sub-sentences, each assigned to final clustered topics (Global connectivity and networking, Information news and learning).}
    \label{fig:pipeline}
\end{figure}

\subsubsection{Topic Discovery}
We treat topics as proxies for ideas: broad thematic areas or conceptual domains that represent new directions of thought introduced into the writing.
To analyze how AI suggestions influenced the ideas expressed in participants' writing, we developed a multi-step topic analysis pipeline.
This pipeline consists of two stages: first, topic discovery, where we identify a set of candidate topics that are neither too granular nor too coarse to label ideas, followed by topic assignment, where these topics are assigned to participant text.

The first step in this pipeline was topic discovery, where we needed to systematically identify relevant topics from participants' text and AI suggestions.
In particular, we needed to be able to assign topics at a granular sub-sentence level, to reflect how participants adapted to new ideas throughout the writing process.

Prior work shows that short snippets are too sparse to support coherent topics, and benefits from expanding them into richer pseudo-documents before modeling \cite{bicalho2017shorttext,li2018shorttext}.
To do so, we employed an LLM-based method to expand out from the sentence-level.
For the topic discovery phase, we used 30\% of the text data from participant writing and AI suggestions, and split them at the sentence-level.
For each sentence, we used \texttt{gpt-3.5-turbo}~\cite{openai-gpt-3.5-turbo} to generate between 0 and 3 relevant topic titles and descriptions.
We also gave \texttt{gpt-3.5-turbo} the previous sentence as context to help the model understand the surrounding text and assign the appropriate topic~\cite{brown2020language}.

Through this process, we obtained several thousand total topics and descriptions, which we then clustered to derive our final topics.

To cluster the topic descriptions together, we employed hierarchical agglomerative clustering into 40 initial clusters based on a \textsc{BERTopic} pipeline for topic clustering~\cite{grootendorst2022bertopic}.
The forty resulting automatic clusters were then manually evaluated using a dendrogram visualization, and merged and split to ensure meaningful groupings that were internally consistent and distinct.
This process was conducted by two researchers, who grouped clusters separately and then discussed them until agreement was reached.
For example, we merged the two clusters ``anxiety'' and ``depression'' together into a single topic called ``mental health,'' since they are closely related.
We manually reviewed a subset of the descriptions within each of the resulting 20 topics to ensure they were accurate and contextually appropriate. 
We include the hierarchical agglomerative clustering visualization and the twenty final topics in the Appendix.

\subsubsection{Topic assignment}
The second step in our pipeline is the labeling of ``writing bursts'' with the 20 topics for further analysis.
We used \texttt{gpt-4o}~\cite{openai2024gpt4ocard} to (1) split the writing bursts into topic-based segments (including uncategorized segments) and (2) assign topics to each topic-based segment (including uncategorized segments). We again used few-shot prompting to help the model understand the task, along with a set of examples of how we wanted the model to label the sentences and negative examples of how we did not want the model to label the segments. We asked the model to split each writing burst into a set of sub-sentences, since one sentence could contain from zero (e.g., generic statements like ``social media is good for society'') to multiple topics (e.g., ``social media can lead to loneliness in teenagers and they could also face cyberbullying'').
To reduce noise, we removed generic terms like ``social media'' and ``benefits'' that appeared frequently but carried little semantic specificity from the essays. 
As a validation to check the contents of the uncategorized segments, we manually labeled a subset of the uncategorized segments ($n=166$) as (1) unclear/incomplete text, (2) general evaluative statements without a specific topic, and (3) specific topics not covered in the 20 identified topics. We found that, based on this manual labeling, any uncategorized content with meaningful topics that were not covered in the 20 identified topics is only 4.7\% of the entire dataset. We decided to remove all uncategorized segments from the dataset.
During prompt refinement, we iteratively manually reviewed subsets of generated labels to ensure they were plausible and contextually appropriate.

\subsubsection{Statistical Analysis}
\label{subsec:stats}
To complement the qualitative findings, we conducted statistical analyses to examine how AI assistance affected participants' writing behavior and content. 

First, we analyzed the time participants spent on their essay based on whether they saw AI suggestions. For statistical analysis, we fitted a linear regression model (formula: time spent $\sim$ treatment group) to evaluate the extent to which the AI treatment predicted participants' time spent on the essay. We also report descriptive statistics of the mean and median time spent on the essay, the percentage of text accepted, and the time suggestions were visible on screen.

Second, we fitted a linear model (formula: topic frequency in text $\sim$ topic frequency in suggestions) to predict relative topic frequency in participants’ final text based on the frequency of a topic in the AI’s suggestions to examine the correlation between topics in the predictions and in participants' writing. We also fitted a complementary model (formula: topic words in text $\sim$ number of suggestions related to topic) based on the number of words participants wrote on a topic compared to how many
related suggestions they saw, as well as a linear model (formula: self-written words on topic $\sim$ number of topic suggestions) predicting the amount of self-written text (as opposed to text accepted from the AI) based on the number of suggestions seen. Finally, we fitted a logistic model (formula: any text written on topic $\sim$ any suggestion seen on topic) to estimate how receiving any suggestions about a topic affected the odds of writing on it. 

%For the relationship between topic frequency in AI suggestions and participants' text, we fitted linear models predicting topic frequency in participants' writing based on the frequency of the topic in the AI's suggestions. For binary outcomes (e.g., whether a topic appeared in a participant's essay), we fitted logistic regression models and report odds ratios with 95\% confidence intervals and p-values.
\section{Qualitative Results: The Reactive Writing Model}
\label{sec:qual_results}

In this section, we summarize the results from our qualitative interviews.
We followed a cue-based retrospective protocol, where participants were asked to reflect on their thoughts at the time of the experiment while they watched the behaviors they adopted.
Through our thematic analysis, we identified three stages that capture the reshaping of the writing process with AI autocomplete suggestions: \textit{attention capture}, \textit{agreement-governed inclusion}, and \textit{post-hoc personalization}.
These three steps become the foundation for what we term \textit{reactive writing}. In the \textit{reactive writing} model, the author is continually redirected to the AI suggestion, pushed to consider whether to accept or reject a suggestion, and finally completes a reconciliation process to adapt the AI text to their own views and experiences. Notably, the AI suggestions influence the writing process regardless of whether the author accepts or agrees with the suggestion. In this section, we describe our qualitative investigations of each of these steps in depth.

\subsection{Attention Capture}
\citet{Kintsch1980DiscourseProduction} suggest that writing starts with an ideation phase that probes writers' long-term memory.
In these models, the writing prompt serves as an initial catalyst to recall pertinent past experiences and knowledge~\cite{flower1981cognitive}.
This ideation-first process starts with memory retrieval and idea generation, followed by composition, and thus assumes writers begin with internal resources and work outward toward expression.
In this section, we describe how participants reported that AI suggestions displace this ideation process and trigger an immediate shift to evaluation.

\subsubsection{Inline automatic suggestions interrupted the ideation process}
As expected from classical writing models~\citep{Kintsch1980DiscourseProduction, flower1981cognitive}, our participants appeared to go through a process of ideation at the very beginning of the writing exercise---they drew on personal anecdotes (P10 described making new friends during the pandemic), observations of close others (P6 recalled how their wife lost friends due to political fights on social media), and broader knowledge from media (P17 referenced reading about the psychological effects of social media). 

Inline suggestions systematically interrupted participants' internal ideation process.
As AI suggestions began to appear on the text editor, participants shared that their attention immediately became focused on the content of the suggestions.
P7 described their initial ideation process, and the experience of watching inline suggestions appear:

\begin{quote}
    \textit{I was going to say something along the lines of social media being very helpful with people with particular interests. You know, the way I use social media is to connect with people that have similar interests to me, and we can advance our knowledge about those topics. So [\ldots] very specific things, broad philosophical debates, trying to get information about certain hobbies that we might have. [\ldots]  And as I started typing, the AI started feeding a full set of this [text] that was along the lines of what I was thinking.}
\end{quote}

In this quote, P7 described how their complex, internal ideation process was redirected to the AI suggestions.

P6 felt these suggestions interrupted their train of thought, saying \emph{``Well, now I'm not thinking about what I'm writing. I'm reading the sentence.''}
Suggestions appeared after 1–2 seconds of pause in typing, and participants felt this interrupted their ideation process.
For example, P1 shared, \emph{``I was just trying to frame a sentence in my mind when I saw the suggestion.''} 
This speed often outpaced participants' own ideation and sentence planning. 
P4 explained, \emph{``[the AI] gave me the suggestion in one second, whereas it would have taken me time to think of something.''} 

\subsubsection{When AI suggestions appeared, participants immediately began to evaluate them, which they attributed to the inline interface design and cognitive tradeoffs.}
Once a suggestion appeared, participants instinctively began to evaluate if they agreed with the suggestion.
P10 explained, \emph{``I was reading the suggestions that the AI provided and I was just thinking—does this agree with how I feel.''} 
Similarly, P8 noted, \emph{``I was just double-checking with myself if I agreed with it. And it made sense.''} 

Since the suggestions were shown inline, participants found them hard to ignore. For example, P8 got the impression that \textit{``I was just getting started, thinking about [the prompt]. And then the suggestions just throw themselves at me.''}
P4 and P6 felt that the inline nature of the text was disruptive to their usual writing process:
\begin{quote}
    \textit{``[The suggestions] made it more difficult for me; I like to write, but I don't like to be hurried, rushed or have words thrown out there that I have to read and they distract me from my thought process. I'm not in my head thinking. I'm reading.'' -- P4}
    \\
    \textit{``I would type, and when I'm typing, I stopped to think and compose my next thought. And then \textbf{bingo}, here comes a sentence. Well, now I'm not thinking about what I'm writing. I'm reading the sentence.'' -- P6}
    
\end{quote}
P6 went so far as to \textit{``type so fast that [the AI assistant wouldn't] throw ideas at me.'' }
Overall, participants felt that the inline format encouraged and sometimes forced them to focus on the AI suggestions rather than their own writing.

Participants also explained that evaluating AI suggestions required less effort---recalling experiences and crafting arguments was more demanding than vetting pre-constructed sentences.
P4 stated, \emph{``It takes more effort for me to come up with it myself than to read it, decide if I agree with it,''} and P8 echoed, \emph{``the whole sentence was already laid out for me. And I didn't really have any issue with it.''} 
Participants explained that the speed of suggestions, and the difficulty of crafting their own sentences, compelled them to think more about evaluating the AI suggestions than producing their own writing.
In these ways, AI suggestions appear ideally suited to interrupt and skew our participants' ideation process towards evaluating the AI suggestions.

\subsection{Agreement-governed inclusion}

When participants evaluated an AI suggestion, they needed to decide whether to accept the text, or reject the text by requesting a new suggestion or typing something else instead.
Our qualitative analysis identified key reasons why writers choose to accept or reject AI-generated suggestions.
Accepting suggestions appeared to be driven by four primary factors: content agreement (alignment with existing beliefs and written text), form appreciation (admiration for polish and eloquence), effort economics (the ease of evaluation versus generation), and default positioning (treating suggestions as starting points to be modified).
Participants appeared readily willing to accept suggestions they tangentially agreed with---potentially suggesting a default effect, where AI suggestions are seen as the most frictionless choice~\citep{dhingra2012default}.
Rejecting suggestions, conversely, was motivated by content disagreement, voice preservation concerns, originality preferences, or perceived disruption.

\subsubsection{Main drivers for accepting and rejecting AI suggestions}

Many participants thought about suggestions as template text that could be modified as needed, which appeared to lower their threshold for acceptance.
P10 expressed: \emph{``For me, suggestions are more like a default.''} 
Because these suggestions were automatically shown in the editor, some participants did not perceive an explicit choice to include or ignore suggestions, but rather thought of these suggestions as the default path forward in writing.
P14, who agreed with a suggestion, explained: \emph{``My thought process was if I agreed with it, I would leave it in.''} 
Conversely, P5, who disagreed, said: \emph{``I think I want to get rid of these suggested texts and to put my own in [...] it is different to my opinion.''} 
In this way, the interface design made accepting suggestions the path of least resistance, with opinion alignment often serving as sufficient justification to retain pre-populated AI suggestions.

In this vein, most participants recalled that their primary reason for accepting suggestions was broad agreement with the content of the suggestion.
P4 explained this common way of thinking: \emph{``I would read the suggestion, decide if I agreed with it, and if I did, I would accept it.''}
Some participants simply accepted suggestions as long as they did not contradict their existing arguments or planned content. 
P7 explained: \emph{``I don't think they hurt any argument that I was making. They just kind of didn't contradict what I was going to say.''}
Overall, participants tended to accept suggestions that matched their overall stance---or did not directly contradict their beliefs.

Participants also often accepted suggestions due to the credible, well-formed prose.
For example, P1 emphatically noted \emph{``the sentences were perfectly constructed!''}
P2 highlighted the AI's efficient communication: \emph{``[The AI] was pretty efficient in what it was saying. It was short, it was succinct, it got to the point.''} 
Participants praised the AI's sentence construction regardless of the content or opinion in the suggestions. 
P14 echoed this perspective, saying, \emph{``the writing is so superior with the assistant, whether it's negative or positive, it's very superior writing.''} 
This admiration for prose led some to ascribe expertise to the AI. 
P1 explained: \emph{``From the vocabulary and how the sentences are framed it sounds like how someone who is learned or knowledgeable in the topic would write; somebody who has researched on it and someone who knows the effects of social media, that's why I kinda started believing in it.''} 
Participants treated the AI's proficiency in language not just as good writing, but as a signifier of authority and expertise (P6: \emph{``I think it came across fairly intelligent.''}) 
For many, the sophisticated language of suggestions appeared to be an important factor in accepting suggestions.

Participants seemed comfortable accepting suggestions they agreed (or did not disagree) with, often regardless of whether they reflected their precise thoughts, ideas, or voice.
As P7 noted: \emph{``they weren't things that I would have written myself specifically,''} or P14: \emph{``there were things that came up that I do agree with, but I hadn't actually thought of at the time that I was writing.''}
P4 candidly expressed: \emph{``I may not have framed it in the same way [...] but I like the way it's written, so I accepted that suggestion.''} 
In these ways, complete alignment between the writer and the AI suggestions was not always necessary to accept suggestions.

Conversely, participants rejected suggestions when they did not cross the minimum threshold of reflecting their voice or experiences.
For instance, P7 rejected a suggestion, \emph{``this suggestion is fine, but I'd rather articulate in my own way what I want to say.''} 
Multiple other participants also rejected suggestions that did not reflect what they knew or felt, for example, P13 recalled \emph{``I don't have any personal experience of that, so I don't think it's a fair comment for me to make.''}, while P9 simply stated \emph{``that was not the thing I experienced.''}
P17, on the other hand, explained that the AI suggestion was not original enough: \emph{``when expressing opinions, I think of how it relates to my life, that's what makes it original.''} They contrasted their own perspective with the AI suggestions, \emph{``[the AI] was giving a very piece-meal average view of social media.''}

Some participants also rejected suggestions they felt were too opinionated or biased.
P13, for example, felt \emph{``the suggestion [...] was more negative than I wanted to be.''}
P5 was particularly dubious of the suggestions: at various points they noted \emph{``I'm noticing how positive these suggestions seem to be [...] I know not to use those''}, and \emph{``it made me feel that the writing suggestion prompt was sort of skewed in its agenda.''} 
Thus, some participants did try to maintain their own ideas or way of writing, despite the AI's encroachment.

\subsubsection{Participants maintained a sense of control and agency over their writing, even when accepting opinionated AI suggestions.}
Throughout the study, participants expressed strong confidence in their agency over the writing. 
Participants felt in control as they were confident that they could reject, accept, or later modify suggestions at any time. 
Some emphasized their power to reject: P14 stated, \emph{``If there was something that I didn't agree with, I wouldn't accept it.''} 
Others felt empowered by their ability to revise, as P2 expressed:
\begin{quote}
    \textit{``I didn't have any pressure, I have my free will, I hope. I didn't feel any pressure as I can always change it. I wasn't hindered by any of the responses and I did not feel I was forced to take any of them. [\ldots]
    I can see that I have deviated from my original neutral stance, but I was fine with it. I always thought that if I disagreed with it, I could take it back.''}
\end{quote} 

This sense of retained agency appeared crucial in participants' comfort with accepting AI suggestions.
Participants perceived the suggestions as easy to change, in part due to the inline and editable interface.
Even P5, who disagreed with the AI's positive stance, chose to accept a few AI suggestions, explaining \textit{``I was exploring how the suggested prompt worked, and thinking about how I would delete it.''}
This retained sense of control may explain why participants felt comfortable accepting suggestions from an opinionated AI assistant.

Despite this sense of control, multiple participants did acknowledge that accepting AI suggestions could lead to unintended shifts in content direction.
P4 was aligned with the AI's negative stance, and said \textit{``I probably would've given a different reason why its harmful.''}
For example, P15, who initially planned to write \emph{``more on the positives''} about social media, ended up accepting critical suggestions because they \emph{``weren't wrong''}: \emph{``Ultimately, no matter how negative the suggestions were, they weren't wrong!''} 
After reviewing their process, P15 acknowledged: \emph{``The suggestions were geared towards fake news, mental health, harassment, all of which we know are real issues!''}. 
These were topics that P15 had not initially planned to address.
Similarly, P1 reflected how \textit{``the suggestion kinda convinced me that the cons are more and my thoughts went in that direction, and I started thinking less about the pros.''}
Thus, participants did write about substantially different content with the AI's help --- and some even noted they had changed their stance.

These shifts in content occurred in spite of (or perhaps because of) participants' sense of control over the suggestions. When asked explicitly whether the AI had influenced their writing, P7 emphasized \emph{``I don't think [I was influenced]. Because in the end I chose not to use some of its suggestions.''}
Yet P2's reflection captures a contradiction: although they recognize their writing deviated from their original stance, they are comfortable with this change since they retained editorial control to accept or reject suggestions.
This broad sense of control may have concealed the AI's true influence on participant writing: the biased selection of ideas.
While participants felt empowered to choose whether to \textit{accept or reject} a suggestion, they could not choose \textit{which} AI ideas they saw.

\subsubsection{Over time, some participants developed trust in the system and grew more comfortable accepting suggestions, while others felt alienated and frustrated}
As participants continued to use the system, positive experiences compounded to increase trust and reliance on the writing assistant.
P1 and P2 both expressed a similar shift in their interaction: \emph{``until the fourth sentence I read the suggestions completely and then accepted them, but then I kinda started believing in it [\ldots] so I started hitting tab when the suggestion appeared without reading the entire sentence.''} 
For participants, initial and recurring agreement with content appeared to build trust in the AI, and participants further softened their conditions for accepting suggestions.
When the AI consistently aligned with participants' writing and stance, they felt very positively about their experience: \textit{``I loved the writing assistant''}, said P14.

Instead of forming a trust-based relationship with the AI, three participants (P5, P6, and P17) experienced the AI primarily as a constant intrusion.
This divergent experience with the AI assistant seemed to stem from misalignment between the AI's stance or writing style, and participants' own.
Overall, P5 felt that the AI did not reflect their own opinions, P17 believed the AI suggestions to be unoriginal, and P6 thought the AI's writing did not match their own style.
Specifically, P5 felt consistently alienated by the AI's positive stance, saying \textit{``I think because it was different to my opinion, it was rather annoying to see them.''}
P6 --- who was a trained writer --- was consistently angered by AI suggestions since \textit{``they almost never were taking my writing where I wanted to go,''} specifically calling out \textit{``I'm going to start with summary and overview and [the AI] jumped right into examples, so that was too early.''}
Overall, alignment or misalignment between the AI and participants' writing compounded into either increasing trust or persistent frustration.

\subsection{Post-hoc personalization}

After participants accepted suggestions, they began the process of adapting the AI text to their own voice, stance, or experiences. This third step of the reactive writing sequence sees writers modifying AI suggestions that they felt were useful, but not a perfect fit. 
We describe how participants cognitively engaged with AI suggestions to edit and elaborate on them, an adaptive process which prior work refers to as \textit{integrative leaps} in a creative writing context~\citep{singh_where_2022}.
We also identify how, even when authors were pushing back or rejecting suggestions, their future writing could be guided by AI-introduced ideas.

\begin{figure*}[htbp]
    \centering
    \includegraphics[width=0.85\textwidth]{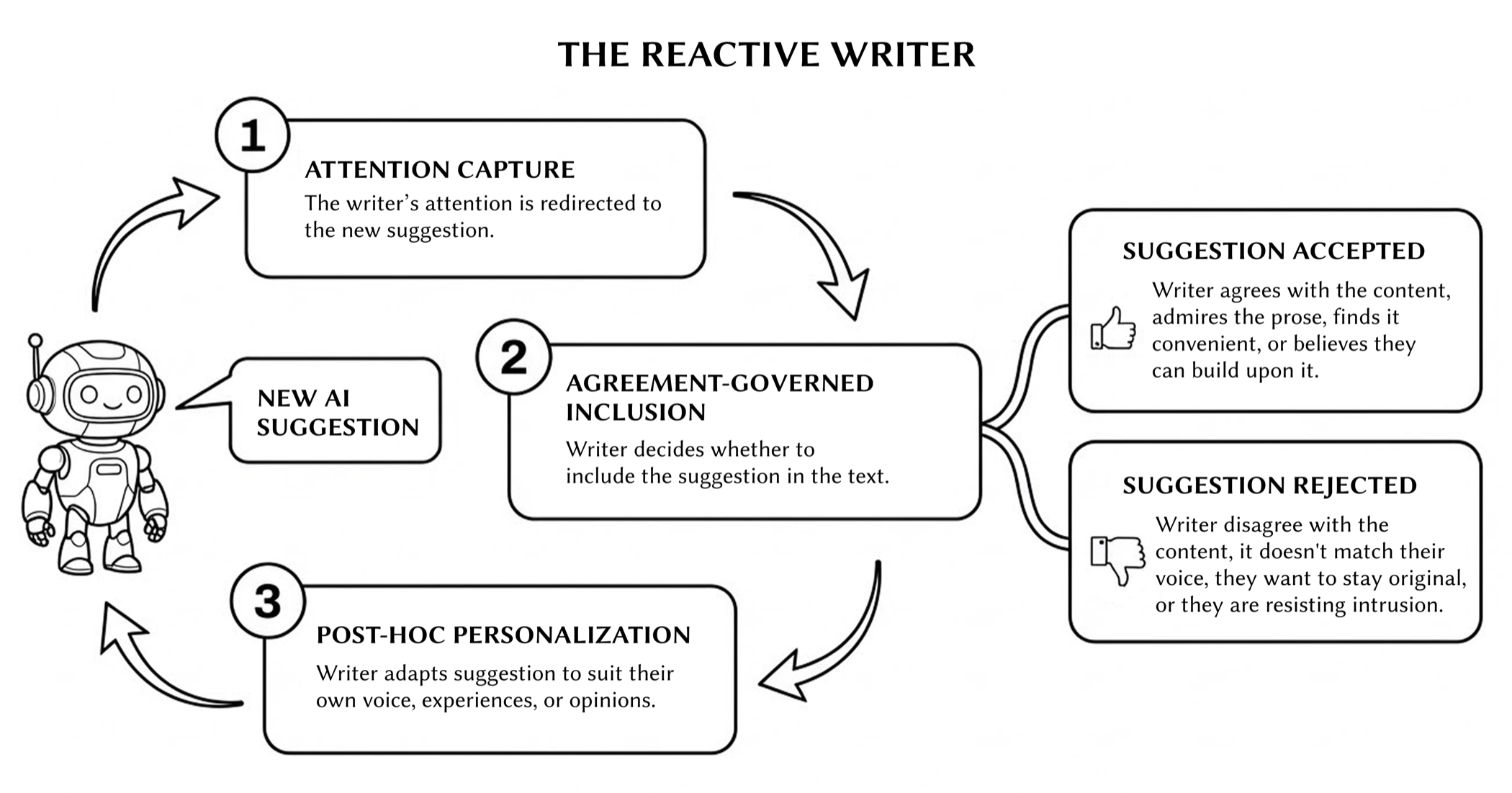}
    \caption{A model of the \textit{Reactive Writing} process. In this model, a new AI suggestion triggers \textit{Step 1: attention capture}, where a writer's attention is redirected from ideation to a new suggestion. In \textit{Step 2: agreement-governed inclusion}, the writer immediately begins to evaluate whether they wish to accept or reject the text in their writing, with a bias towards accepting suggestions. The diagram lists the main reasons to include or reject the text. When text has been accepted, writers turn to \textit{Step 3: post-hoc personalization}, where they incorporate and adapt the suggestions to their own writing.}
    \label{fig:reactive-writer}
    \Description{A flowchart showing the Reactive Writer model with three numbered steps: (1) Attention Capture, (2) Agreement-Governed Inclusion, and (3) Post-hoc Personalization. A robot icon represents the AI generating new suggestions. Two boxes on the right show reasons for accepting suggestions (agrees with content, admires prose, finds it convenient, believes they can build upon it) and rejecting suggestions (disagrees with content, doesn't match voice, wants to stay original, resisting intrusion).}
\end{figure*}

Notably, our findings also highlight that AI suggestions influence the writing process regardless of whether the writer chooses to accept a suggestion.
For example, if the AI suggestion presents a new idea that the writer chose not to accept, the suggestion may still prompt them to think about related ideas or to write a counter-argument.
Even when not extensively accepting suggestions, writers could still be reacting to the AI suggestion and engaging in post-hoc personalization, rather than engaging in their own cognitive processes that trigger novel ideas. 
There were writers in our study who resisted this shift and attempted to maintain a generation-first process, but they often struggled against the interface's inline suggestion design.

\subsubsection{Participants made edits to the AI suggestions to preserve their own voice, avoid copying, and improve the phrasing, but often preserved underlying framing.}
Participants made edits to accepted suggestions for a variety of reasons.
P2 explained that they made edits to retain their own voice: \emph{``Maybe shorten [the suggestions] a little bit. This uses my own words that I use more often just to make it a little more personal.''}
P15 felt that using the AI's suggestions verbatim felt like ``copying:'' \emph{``I can't copy just from, copy straight from there. I have to be in my own kind of thing.''}
And for concision, P3 explained, \emph{``I just wanted to make it a little bit shorter.''}
In these ways, even when suggestions were kept as conceptual placeholders or defaults, participants reshaped them to sound like themselves, fit their own experiences, and read more naturally.

Often, when participants edited the AI suggestions after accepting them, they preserved the central framing of the suggestion.
The simplest edits were superficial stylistic adjustments, such as replacing ``however'' with ``but,'' or adjusting verb tenses—changes that preserved both meaning and structure. 
More substantive edits involved fine-tuning content to align with personal views while remaining anchored to the suggestion's core idea.
For instance, P7 removed ``overall'' from the AI-suggested phrase ``overall, social media is good for society,'' and explained: \emph{``I was thinking whether 'overall' was a good word to include because, you know, I am not so sure whether the positives outweigh the negatives.''} 
While this edit shifted the meaning from a definitive to a more qualified stance, P7 was still working within the AI's initial framing. 
Even participants who extensively rephrased suggestions acknowledged they often preserved the underlying concepts. As P15 succinctly put it: \emph{``It's the same ideas, but just written differently.''}

\subsubsection{Participants actively related AI ideas back to their own experiences or perspectives, including when they disagreed with the suggestion.}
After participants accepted an AI suggestion, they often engaged in additional cognitive work to personalize the suggestion.
P7, for example, sought to make the accepted AI suggestions more concrete by describing ``how'' and ``why'' they agreed with the AI text: 
\begin{quote}
    \emph{``I felt that I needed to be a little more specific to share my feelings because both sentences [the AI] provided for me were a little generic. I agreed with them, but I wanted to specify 'how' [social media] connects people from all over the world. [...] Because I kind of accepted this sentence that was generated for me, I wanted to elaborate on 'why' I said it. And that's when I went into writing about my own personal experience.''}
\end{quote}
Participants would also connect ideas back to their own life and experiences.
P2 shared, \emph{``You know, as this popped up, I started to think about how I use [social media], even during work, to raise awareness about certain issues.''} 
P8 similarly described, \emph{``I was basically just expanding on the point using more concrete examples that I've seen before.''}
Through personalizing text suggestions, participants made the AI ideas their own by combining the suggested text with their own world views and experiences.
Even after rejecting a suggestion, P9 stated \emph{``I was just thinking of how I saw the suggestion before, and then I was wanting to put my spin on.''}

Beyond directly adapting the text, AI suggestions appeared to function as ideation cues that redirected participants' flow of thought. 
P18 noted, \emph{``As soon as the suggestion came up, in my mind, I was like: Oh, yeah, that makes complete sense! And my mind went in that direction.''} 
For some, the assistant seeded topics they were not previously considering: P16 reflected, \emph{``I wouldn't have thought of these things like 'misinformation' or 'hate speech' by myself. Suggestions help me kickstart that thought.''} Thus, the AI suggestions seem to trigger new ideas for participants.

Even in those cases when participants did not agree with AI suggestions, the suggestions still influenced their writing by prompting counterarguments.
For example, P5 faced an AI with a much more positive view of social media than them.
Rather than simply ignoring the suggestions, P5 began to counter the points the AI was making in the writing.
As the AI introduced new ideas, P5 decided \emph{``I think that I want to get rid of these suggested texts and to put my own in, which is much more negative.''}
P5 felt the positive suggestions were \emph{``making me want to argue against this [point], because it disagrees with my line.''}
P12 had a similar experience when they saw a suggestion they disagreed with and wrote about the same topic but in a negative light: \textit{``[the suggestion] was different to my opinion, it made me want to push back at it.''}
This example demonstrates how AI-introduced ideas can influence writing even when rejected, prompting writers to articulate opposing arguments.

Overall, participants reported that AI suggestions reshaped their writing process, shifting attention away from ideation and toward evaluating pre-generated text. They often accepted suggestions when they agreed with the content or admired the fluent prose, even if it did not fully reflect their original ideas or voice. At the same time, participants frequently edited or adapted accepted suggestions to align better with their perspectives, maintaining their sense of control over their writing process even though the AI system had cued the idea generation underlying what they wrote about.

\subsection{The Reactive Writing Process}

Our findings suggest that AI-assisted writing reshapes the cognitive writing sequence.
Among our participants, evaluation of suggested text appeared to displace personal ideation as the primary cognitive activity in early composition.
Writers shift from asking \emph{\enquote{What do I think about this?}} to \emph{\enquote{Do I agree with this suggestion?}}
Co-writing with AI seemed to be fundamentally reorienting their writing process: instead of generating content from memory, many writers generated impulses for their writing process by deciding whether to adopt, reject, or adapt external content.

We call this process \textbf{reactive writing:} writing where authors ``ideate'' by reflecting on external content, rather than self-generated ideas.
In reactive engagement with suggested ideas, the entry point to composition becomes a judgment of agreement with the suggestion, rather than memory retrieval. 
Personal experiences, examples, and voice are subsequently used to respond to and elaborate on suggested claims rather than as initiators of new directions.
Figure~\ref{fig:reactive-writer} visualizes this process, which has three key characteristics.
First, AI suggestions lead to \textbf{attention capture}, where participants' attention is redirected from their own ideas to the suggestion presented.
Second, writers engage in \textbf{agreement-governed inclusion}, where fit with existing beliefs or written text becomes the primary filter for incorporating material.
Third, writers conduct \textbf{post-hoc personalization}, where individual voice, experience, and opinions serve to elaborate rather than originate the conceptual direction.
Within the reactive writing process, the role of a writer shifts from being primarily a generator of content and ideas to being an evaluator and extender who judges, adapts, and contextually elaborates AI-generated content.

\section{Exploratory Quantitative Results}
\label{sec:quant_results}

In the following sections, we explore the intuitions provided by the qualitative results and \textit{reactive writer} model in a quantitative analysis of 1{,}291 co-writing sessions.
We focus the quantitative analysis on three aspects of the reactive writing process. First, we examine attention capture through the lens of session duration (Section 5.1). Second, we examine the downstream effects of agreement-governed inclusion and post-hoc personalization by analyzing the topic distributions in the AI's suggestions and participants' texts. We assess whether the topics suggested by the model correlate with the topics prevalent in the final text (Section 5.2), and explore the extent to which the suggestions may have influenced participants' writing through cognitive engagement beyond the mere acceptance of text (Section 5.3). 

\begin{figure}
  \begin{center}
    \includegraphics[width=.33\textwidth, trim=0cm 0.7cm 0cm 0, clip]{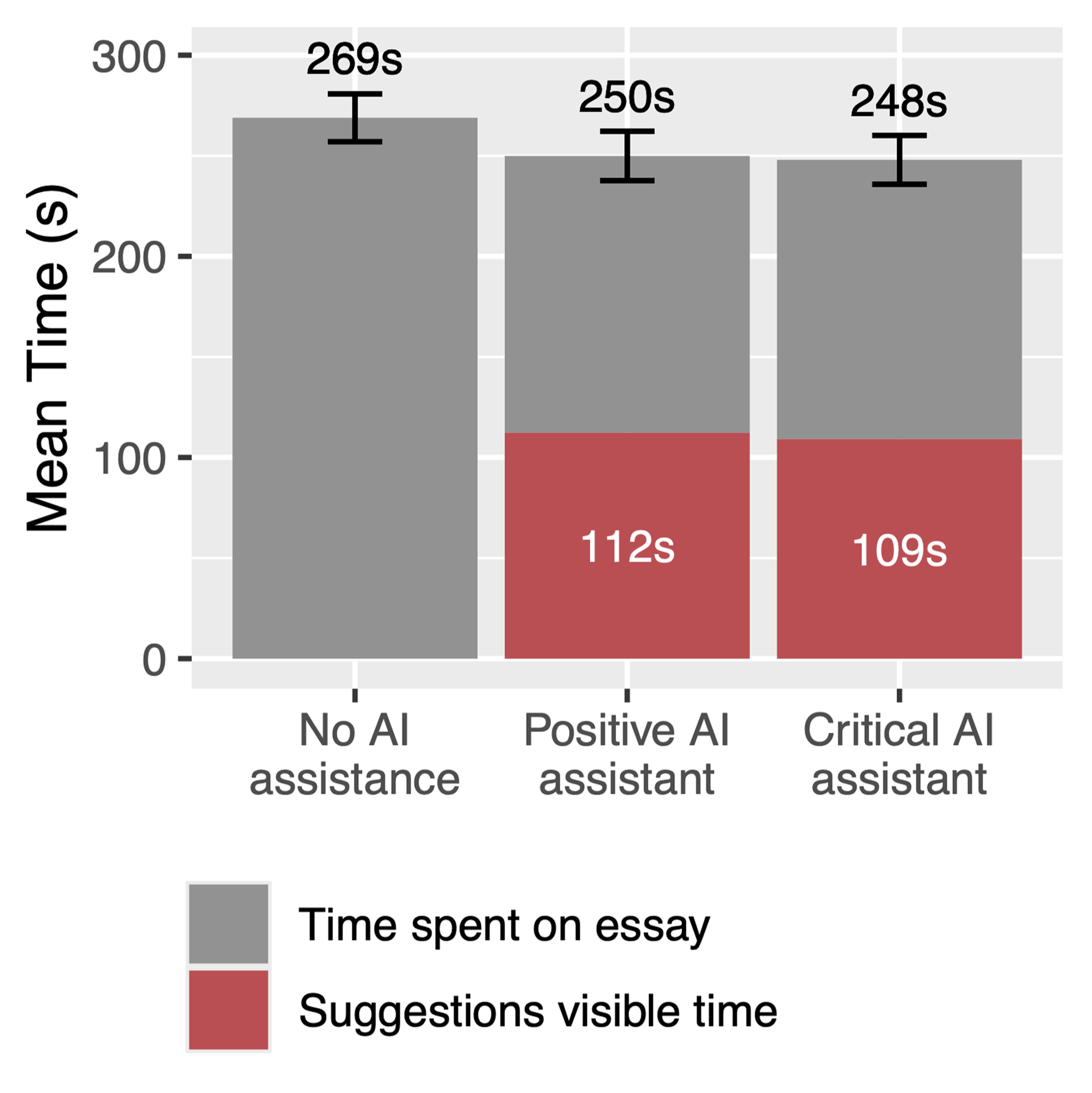}
  \end{center}
  \caption{\textbf{Writing with AI did not substantially reduce the time participants spent on the essay.} \textit{Mean task time with 95\% confidence intervals in N = 1,291 writing sessions.} When writing without AI assistance, participants, on average, wrote for 269 seconds. Participants writing with AI assistance spent about 248-250 seconds on their essay.}
\Description{Bar chart comparing mean task time across three conditions: Control (no AI) at approximately 269 seconds, Positive AI at approximately 250 seconds, and Critical AI at approximately 248 seconds. Error bars show 95\% confidence intervals. The differences between AI conditions and control are small.}
\label{fig:time}
\end{figure}

\begin{figure*}
  \begin{center}
    \includegraphics[width=.9\textwidth, trim=0.25cm 0 0.25cm 0, clip]{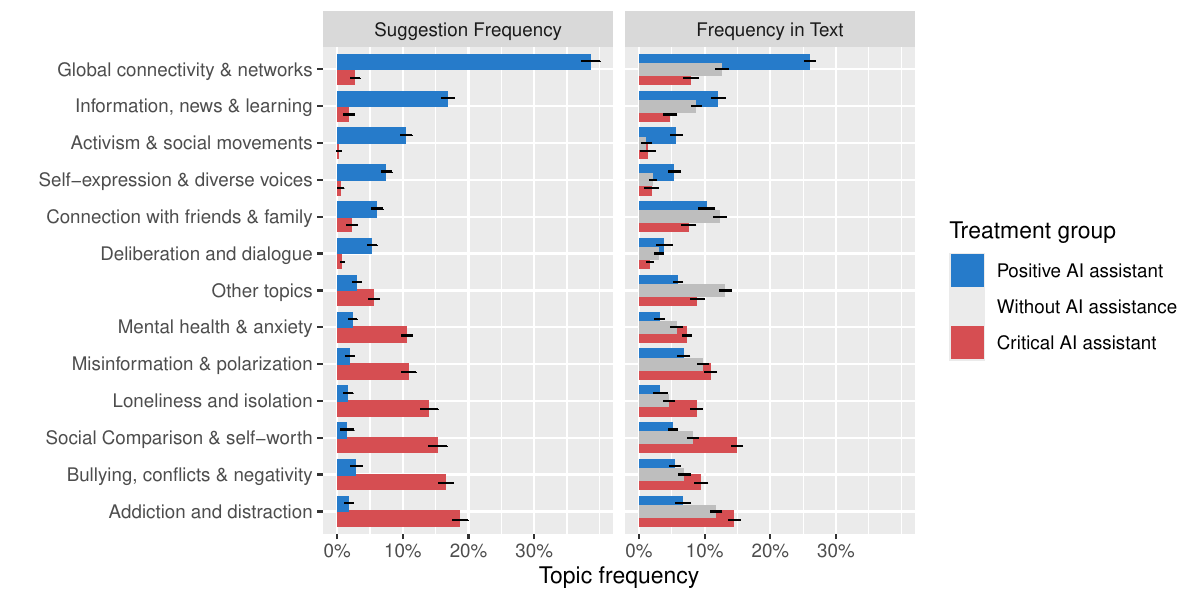}
  \end{center}
\caption{\textbf{Frequency of topics in the AI suggestions (left) and final text (right) when writing with a \textcolor{blue}{positive} or \textcolor{red}{critical} AI assistant.} \textit{Grey bars indicate control group frequencies for reference. N=1,291, 95\% confidence intervals shown in black.} The AI assistant's suggestions (left panel) showed a clear tendency towards certain topics, with many suggestions related to global connectivity in the positive AI treatment group, and many suggestions related to addiction, bullying, and loneliness in the critical AI treatment group. Participants' written text (right panel) inherited the topical biases of the suggestions to a substantial extent.
}
\Description{Bar charts comparing topic frequencies in AI suggestions (left panel) and in final essays (right panel) for positive (blue) and critical (red) assistants, with grey bars marking control group frequencies. The left panel shows assistants emphasize stance-congruent topics; the right panel shows these biases partially carry over into participants' writing.}
\label{fig:distribution}
\end{figure*}

\subsection{Attention capture: Time spent on essay} 
We start by examining signs of attention capture through the lens of interaction time. Our qualitative analysis highlighted how the AI assistant captured participants' attention, outpacing their own writing process. To make use of the AI's suggestions, participants needed to evaluate them. The time participants spent on evaluating suggested ideas not only took away from their personal idea generation process, but also limited the time savings participants could realize through using the AI assistant.

We find that, on average, participants writing with AI assistance accepted about 31\% of their essay text from the AI. 
However, their time savings realized were substantially less than 31\%, as shown in Figure~\ref{fig:time}: Participants in the control group spent 269 seconds on their essay when writing without AI. Participants writing with AI assistance, in contrast, spent 250 or 248 seconds. We fitted a linear model to predict time on task based on treatment group. The time differences in both the positive AI treatment ($\beta = -18.98$, 95\% CI [-36.05, -1.91], t(1512) = -2.18, p = 0.029)  and the critical AI treatment ($\beta = -20.91$, 95\% CI [-38.11, -3.71], t(1512) = -2.38, p = 0.017) are statistically significant, but correspond to time savings of only 7.5\%.

In addition, we evaluated the time difference for the median participant as an alternative estimate less affected by participants who heavily relied on the AI or did not use it at all. The median participant in the treatment groups accepted 28 words from the AI's suggestions in their 112-word essay, that is, about 25\% of their writing came from the AI. At the same time, the median participant in the treatment group, writing with AI, spent about 213 seconds on their essay, compared to the median participant in the control group, who finished their essay in 237 seconds. The difference of about 24 seconds corresponds to time savings of only about 10\% when writing with AI. 

At the same time, treatment group participants spent a substantial amount of time (109 to 112 seconds) with suggestions shown on their screen---a period in which they were not writing, as suggestions appeared only after about 2 seconds of inactivity and disappeared once participants resumed writing. While participants in the control group naturally were not writing without interruption either, a simulation of hypothetical suggestions in the control group shows that they would have appeared for only 72 seconds, suggesting that, at a minimum, the remaining 37-42 seconds were spent evaluating suggestions. This remains a conservative estimate, as participants writing with AI may have also evaluated suggestions at times when participants in the control group paused their writing to generate ideas and plan next steps.

Taken together, these observations align with the qualitative results, suggesting that participants spent substantial time and effort to make use of the AI's suggestions, which not only limited the time savings they realized through using AI but also likely took away from other parts of their writing and ideation process.

\subsection{Agreement-governed inclusion: How much did suggestions affect the topics discussed?} 
In the interviews, we found that many participants were willing to accept suggestions they agreed with. Agreement-governed inclusion could have different effects at scale depending on how strict participants' threshold for inclusion is: strict agreement might mean that the incorporated text largely reflects participants' views, keeping the writing close to original intentions. Alternatively, a low threshold for agreement where participants agreed with a wider range of suggestions would allow the model's suggestions to substantially steer the direction of participants' writing. To estimate the influence that the suggestions had on participants' writing, we conducted a topic-based analysis of the AI's suggestions and participants' essay texts to estimate the cumulative effects on what people write.

In the analysis, we split the interaction logs of 1{,}291 co-writing sessions into units of writing. These writing bursts were then matched to the topics they discussed (see Section 3.2 for details). 
An overview of the resulting topics and frequencies is shown in Figure~\ref{fig:distribution}: In the left panel, the figure shows how often a topic surfaced in the AI's suggestions, while the right panel shows the topic's share of text in participants' essays. Blue bars correspond to participants writing with a positive AI assistant, while red bars show topic frequencies for participants writing with the critical assistant. Grey bars indicate the control group frequencies written without AI suggestions as a benchmark.

The left panel of Figure~\ref{fig:distribution} confirms the intuition from the qualitative interviews that the writing assistants predominantly suggested a specific set of topics:
Participants writing with the positive assistant received proportionally more suggestions about ideas that convey the potential benefits of social media, such as ``Global connectivity \& networks'' and ``Information, news \& learning.''
In contrast, participants using the critical assistant received more suggestions about harms—such as ``Addiction and distraction'' and ``Bullying, conflicts \& negativity.'' 
This figure also shows that ideas likely fall on a stance-congruence spectrum: for example, the ``Connection with friends \& family'' topic is introduced by both the positive and critical AI assistants, suggesting that there can be both positive and negative ways to frame certain topics.

The topical biases of the AI assistant's suggestions translated, although imperfectly, into the topical bias observed in participants' final text, shown in the right panel. 
Relative to the control benchmark, the positive assistant group devoted more essay text to topics aligned with positive topics (for example, social media benefits like ``Global connectivity \& networks'' and ``Connection with friends \& family'').
On the other hand, the critical assistant group devoted more text to negatively aligned topics (for example, social media harms like ``Addiction and distraction,'' ``Bullying, conflicts \& negativity,'' and ``Loneliness and isolation''). 
The observed pattern is consistent with the idea that mere agreement might have posed a relatively low threshold for acceptance, allowing the AI assistant to substantially shape the topics participants wrote about. A statistical analysis of the relationship is provided in Section 5.3.

\begin{figure}
  \begin{center}
    \includegraphics[width=.5\textwidth, trim=0 0 0 0, clip]{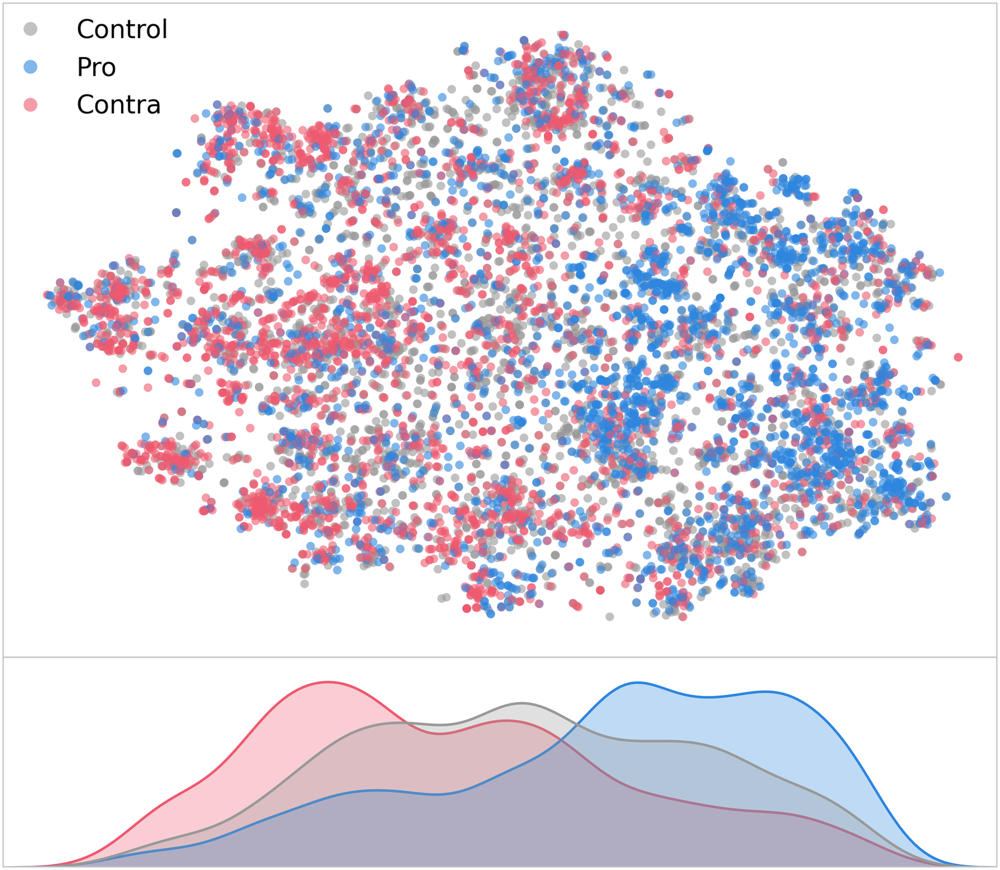}
  \end{center}
\caption{\textbf{Embeddings of participants' sentences when writing with a \textcolor{red}{critical (red)} and a \textcolor{blue}{positive AI assistant (blue)} compared to those writing \textcolor{gray}{without AI assistance (grey)}}. \textit{t‑SNE embeddings of all sentences in N=1,290 writing sessions.} Participants writing with AI assistance wrote about a narrower and systematically different set of topics, while those writing without AI assistance covered a broader area of topics. The marginal density distribution (bottom) shows the horizontal separation of groups along the primary dimension, with critical AI (red) concentrated on the left, positive AI (blue) on the right, and control (grey) distributed across the full range.}
\Description{t-SNE scatter plot of sentence-level embeddings across conditions: red points (critical AI) and blue points (positive AI) cluster into distinct regions on opposite sides of the plot and cover a narrower area than grey points (no-AI control), which are distributed across the full range, indicating more constrained and systematically different topic coverage when writing with AI.}
\label{fig:trajectories}
\end{figure}

Figure~\ref{fig:trajectories} provides a more graphical overview of the shift in topics discussed when participants were receiving positive or critical AI writing suggestions. Using the same topic pipeline, we generated sentence‑level embeddings for each text segment in the control and treatment datasets via the OpenAI embeddings API. We projected the resulting embedding vectors to two dimensions using t‑SNE~\cite{maaten2008visualizing} and visualized them on a scatter plot where points are colored based on whether participants wrote with a critical AI assistant (red), a positive AI assistant (blue), or without AI assistance (grey). The graph shows a systematic separation between topics discussed by participants writing with a critical and positive AI assistant, corresponding to the observation that red points cluster around topics on the left while blue data points cluster around topics on the right. It also shows that the breadth of topics discussed in the treatment groups is narrower for participants writing with AI: The topics participants discussed when writing with AI are clustered in the left (red) or right (blue) side of the graph, while the topics participants discussed when writing without AI are distributed more broadly over the entire graph (grey). This observation also aligns with the idea that, while participants felt in control of their writing, their relatively low threshold for agreement allowed the AI to substantially change the topics they wrote about.

\begin{figure*}
  \begin{center}
    \includegraphics[width=.9\textwidth, trim=0 0 0 0, clip]{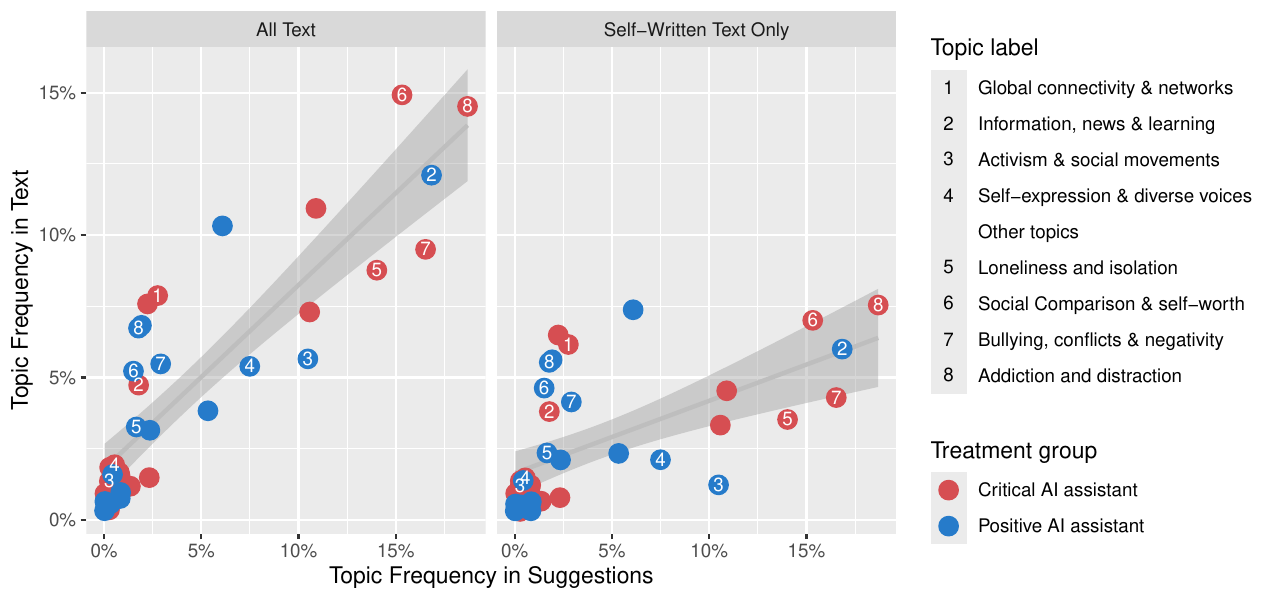}
  \end{center}
\caption{\textbf{Correlation between how often the AI suggested a topic (x-axis) and how much participants wrote about it (y-axis)} \textit{Topic correlations with a fitted linear regression line and 95\% confidence bands. N=1,291.} The left panel shows topic frequencies in the overall text, while the right panel shows topic frequencies in self-written text only. Topics that were suggested more often by the AI were both more likely to appear in participants' final text and their own writing.
}
\Description{Topic correlations with a fitted linear regression line and 95\% confidence bands. N=1,291. The left panel shows topic frequencies in the overall text while the right panel shows topic frequencies in self-written text only. Topics that were suggested more often by the AI were both more likely to appear in participants' final text and their own writing.}
\label{fig:correlation}
\end{figure*}

\subsection{Post-hoc personalization: Influence beyond accepted text}
In the interviews, some participants discussed how they actively engaged with the suggestions, expanding on them with their own experience, writing them in their own words, or responding in other ways. This engagement with the AI's suggestions would suggest that we might see indirect influence in the topic data, i.e., signs of the suggested topics appearing in participants' writing beyond accepted text. The remainder of this section explores the correlation between the topics suggested by the AI and the topics that we observe in participants' writing---both in the overall text and the text they wrote themselves.

Figure~\ref{fig:correlation} shows the correlation between how often the AI suggested a topic (x-axis) and how often the topic occurred in participants' text (y-axis). 
The left panel shows the frequency of topics in the final text, while the right panel shows how much participants themselves wrote on the topic. Each topic is shown by a point, for example, the red (8) in the top-right indicates that the topic ``Addiction and distraction'' accounted for 18.6\% of the suggestions in the critical AI assistance treatment, and the topic also showed up in 14.5\% of participants' final texts. The corresponding point in the right panel shows that the text users wrote themselves on ``Addiction and distraction'' made up 7.54\% of their text.

We fitted a linear model to predict topic frequency in participants' final text based on the frequency of a topic in the AI's suggestions (compare left panel). The model explains a statistically significant and substantial proportion of variance ($R^2$ = 0.85, F(1, 38) = 212.06, p < .001, adj. $R^2$ = 0.84). The correlation between topic frequency in suggestions and final text is statistically significant and positive ($\beta$ = 0.64, 95\% CI [0.55, 0.72], t(38) = 14.56, p < .001). We fitted a similar model predicting how much people would write on a topic based on how frequently the topic appeared in the suggestions (compare right panel). The model explains a statistically significant and substantial proportion
of variance ($R^2$ = 0.51, F(1, 38) = 40.11, p < .001, adj. $R^2$ = 0.50) and indicates a statistically significant and positive correlation ($\beta$ = 0.24, 95\% CI [0.17, 0.32],
t(38) = 6.33, p < .001; Std. beta = 0.72).

We also fitted a complementary set of models based on the number of words participants wrote on a topic compared to how many related suggestions they saw (as opposed to the relative frequencies analyzed above). A model predicting the number of words written on a topic based on how many suggestions a participant saw on the topic indicates a statistically significant and positive correlation corresponding to about 4.43 words in the final essay per suggestion on the topic ($\beta = 4.43$, 95\% CI [4.14, 4.72], $t(16779) = 29.75$, $p < .001$). 
We conducted a similar analysis on the self-written text only: Here, a model regressing how many words a participant would write on a topic (discounting text accepted from AI) on the number of suggestions seen on the topic predicts an additional 1.95 writer‑produced words per suggestion on the topic ($\beta = 1.95$, 95\% CI [1.79, 2.11], $t(16778) = 13.30$, $p < .001$).

We conclude with a logistic model that predicts whether a participant would write any text on a topic based on whether they saw any suggestions on a topic. The fitted model predicts that seeing at least one suggestion on a topic increases the odds that the topic appears in participants' essays by a factor of 3.97 ($\mathrm{OR} = 3.97$, 95\% CI [3.65, 4.23], $p < .001$). When predicting whether participants would write any text on the topic themselves, the odds remain elevated ($\mathrm{OR} = 1.65$, 95\% CI [1.50, 1.82], $p < .001$). 

Our finding that the AI's influence on the topics discussed extended even to text participants wrote by themselves is consistent with participants' qualitative descriptions of elaborating on AI‑proposed ideas and using suggestions as ideation cues.
Taken together, the quantitative analysis provides a large-scale perspective on changes in participants' writing that align with the qualitative observations that the AI assistant interfered with participants' ideation process, and substantially shaped the topics and ideas they wrote about.

\section{Discussion}

In this mixed-methods work, we provide empirical evidence for how inline AI writing assistance may reshape writing processes and text outputs. Through analysis of qualitative interviews, where participants interacted with their own writing using a custom replay tool, we arrive at a new model for AI-assisted writing, \textit{reactive writing}. The quantitative analysis supplies further evidence for this model, validating that participants spent a significant portion of time evaluating external suggestions, and that AI-generated ideas permeate AI-assisted text, both directly and indirectly.

First, our findings on \textit{attention capture} show how inline suggestions can disrupt the traditional internally-generated writing ideation process.
Classical models propose that writing begins with a prompt, which stimulates internal ideation and memory retrieval processes, before writers translate these internal processes into text~\cite{Kintsch1980DiscourseProduction, flower1981cognitive}.
In reactive writing, this prompt is no longer static, but instead continuously updates through new AI suggestions.
Participants in our study reported that nascent ideas were repeatedly interrupted by AI suggestions: moments of pausing to think were co-opted by new suggestions, shifting participants' attention to reading and evaluating them instead.
Our findings suggest that reactive writing, therefore, takes place alongside the classical writing process.
This interruption pattern is unsurprising given that research on digital interruptions finds that even short, task-relevant alerts fragment ongoing cognition~\cite{mark2008cost,iqbal2007disruption}.

Second, \textit{agreement-governed inclusion} helps explain why suggested ideas, once surfaced by the AI, so often enter the participants' draft.
Participants in our study exhibited a bias for accepting suggestions, even when these ideas did not quite match their own voice or thought processes.
There are many known cognitive biases and effects that may explain this tendency.
For example, the tendency to accept AI suggestions could reflect a default effect, where preselected options exert disproportionate influence through cognitive ease~\cite{johnson2003defaults}, and automation bias, where people overweight automated recommendations~\cite{skitka1999automation}.
In some ways, participants appeared to be \textit{satisficing:} once a suggestion roughly matched their opinion and read fluently, accepting the text was easier than coming up with or searching for an alternative~\cite{artinger2022, simon1956rational}.
Fluency and polish of AI-generated sentences further encouraged acceptance, aligning with work on processing fluency, where ease of processing is reliably associated with more positive evaluations and perceived quality~\cite{reber2004processing}.
Despite the ready adoption of AI suggestions, participants maintained that they felt fully in control, noting that they could always revise or reject suggestions.
Prior work on smart-reply systems similarly argues that time pressure and priming can transfer some agency to the AI, even when users retain control over their choice to accept or reject the suggestion~\cite{wenker2022who}.

Third, \textit{post-hoc personalization} reveals how AI-originated content becomes enmeshed with personal voice and experience.
Participants often edited accepted suggestions to sound more like them, shorten phrasing, or add specific examples, yet typically preserved the underlying framing and main ideas introduced by the AI.
This reflects prior observations that writers use AI output as a semantic scaffold and then perform \textit{integrative leaps}, doing cognitive work to make suggestions useful to their own narratives. While~\citet{singh_where_2022} identified these leaps in a creative writing task, our findings show that these leaps also happen when writers are representing their own opinions.
This adaptation allows writers to maintain a sense of authorship by shaping the AI's framing to fit their perspective, even as that framing remains largely intact. In turn, the AI ideas gain legitimacy and stability within the writer's text, which may partially explain how persuasion effects emerge.
This reactive pattern is consistent with classic priming research, which shows that even brief exposure to particular concepts can increase their mental accessibility and bias how people interpret and remember subsequent information~\cite{higgins1977category}. 
Through the lens of cognitive offloading~\cite{risko2016cognitiveoffloading,sparrow2011google,clark1998extendedmind}, reactive writing redistributes cognitive labor: writers remain engaged in evaluation and editing, but their ideas are scaffolded by the model rather than generated independently.

\subsection{Implications for Persuasion}
Beyond these cognitive mechanisms, our findings suggest that reactive writing creates conditions under which biased suggestions can shape writers' expressed and actual opinions. Prior work has shown that autocomplete systems can shift both the content people write and the attitudes they later report~\cite{jakesch2023co,williams2025biased}, but the processes underlying these effects have remained unclear.
The mechanisms we identify may offer a process-level explanation for persuasion in inline predictive writing scenarios. First, the interface creates the conditions for writer attention to be directed to the AI suggestions. Second, cognitive heuristics (e.g., default effects and automation bias), coupled with perceived retained agency, lower the bar for accepting AI-generated text. Third, personalizing the AI text embeds the AI's framing inside the writer's own tone, experiences, and reasoning.
This dynamic aligns with theories of self-persuasion, which show that people are more influenced by arguments they generate or elaborate on themselves than by those delivered by an outside source, especially when they feel high agency over their decisions~\cite{damen2015re}. 
Recent work on conversational agents emphasizes persuasion arising from extended, deliberative exchanges with chat-based systems~\cite{salvi2025conversational,costello2024durably}. Our findings may point to a complementary pathway in which influence emerges as writers elaborate on AI-suggested text within the flow of composition.
While our work is not strictly causal, our findings suggest the introduction of systematically biased ideas coupled with self-reflection and cognitive engagement as promising directions for causal investigation in future work on AI persuasion in autocomplete scenarios.

Our findings also resonate with agenda-setting theories in mass communication. Classic agenda-setting research shows that the issues made salient by news organizations shape which topics the public comes to view as important~\cite{mccombs1993evolution,mccombs1972agenda}. 
As Bernard Cohen famously summarized, the press ``may not be successful in telling people what to think, but it is stunningly successful in telling people what to think about''~\cite{cohen2015press}. 
In our study, AI suggestions played a similar role: by surfacing particular framings at key moments of ideation, they anchored participants' thoughts around those framings and influenced what ideas felt natural to pursue next. Participants only accepted ideas they agreed with, but because the set of ideas they encountered was filtered through the opinionated assistant, they could only evaluate those that the AI chose to present. This mechanism maps onto contemporary theories of algorithmic agenda-setting, where algorithms embedded in everyday technologies shape the informational environment by selectively foregrounding certain content~\cite{trielli2022algorithmic,einarsson2025algorithmic}.
Emerging work has begun to wrestle with the idea that language models may also function as algorithmic agenda-setters~\cite{sichach2024mainstream}, and our work substantiates this theory with empirical evidence.

Our findings and related discussion should be understood in the context of an opinionated AI assistant that introduced stance-laden suggestions into the writing process. While certain dynamics such as attention capture and satisficing may extend to more `neutral' predictive text tools, the persuasive effects we document may depend on writers engaging with an opinionated AI assistant. Therefore, the strongest implications concern systems that offer opinionated suggestions.

\subsection{Design Implications}
Through the reactive writing model, our findings highlight concrete ways in which design decisions can shape both the process and outcome of writing. In particular, patterns such as attention capture, interruption of internally generated ideas, and default-like acceptance of pre-populated text arise from the inline suggestion paradigm itself, and thus point directly to where designers should rely on inline completions versus where they should withhold or relocate suggestions.
A key implication is that assistants should distinguish between speeding up expression of existing ideas and providing new ideation.
Frequent inline completions are particularly potent interruptions precisely because they are rapid and spatially near the cursor; designers could reserve inline suggestions for moments when the writer's direction is clear, so completions reinforce rather than redirect their thinking.
In contrast, suggestions with greater potential to shift topic, stance, or framing might appear less frequently and in more peripheral locations, such as a sidebar, where writers can engage with them deliberately. 
Finally, presenting a single polished completion can make that option feel like the default path forward; offering alternative options and using interface cues to show how each candidate would change the surrounding text can clarify the underlying choice and make the writer aware of undue AI influence.

\subsection{Limitations}
While our mixed-methods design provides rich insights into both the process and outcomes of AI-assisted writing, several limitations should be noted. First, our experimental setting may not fully capture the range of contexts in which writers use AI tools. The salience of opinionated suggestions in our setup could also amplify effects that may be subtler in real-world systems, limiting ecological validity. Additionally, our qualitative sample was recruited exclusively from US- and UK-based Prolific participants, which limits professional and cultural diversity. Further study is needed to investigate the extent to which these findings generalize to other types of writing contexts, interaction designs, and populations. Second, our findings reflect a deliberately strong form of reactive writing that combines opinionated language models with frequent inline autocomplete suggestions. These design choices foreground frictions and disagreements that make underlying mechanisms visible, but may overstate their prevalence relative to more `neutral' models or less intrusive paradigms such as chat-based planning tools. Future work should examine whether similar dynamics arise when assistance is less stance-laden, appears less frequently, or is presented in more temporally and spatially distant formats.

\section{Generative AI Disclosure}
We acknowledge the use of generative AI tools to improve research efficiency and writing clarity in this manuscript. We used these tools as follows: ChatGPT was used for literature discovery through search assistance to identify relevant research papers for the related works and discussion sections (alongside traditional tools like Google Scholar and existing Zotero libraries). Claude and ChatGPT were used for refining and streamlining manuscript content, as well as specific formatting tasks such as translating the interview protocol to LaTeX. All manuscript text was written and finalized by the authors.

%% Maurice often doesn't do conclusions, but we should have acknowledgements here. Note that we also need to acknowledge potential LLM use in writing.

% commenting out for anonymous manuscript
% \begin{acks}
% We thank Yujie Shao for her support in performing the pilot studies for this project. This material is based upon work supported by the National Science Foundation under Grant No. CHS 1901151/1901329 and the German National Academic Foundation. Part of the work is funded by the Bavarian State Ministry of Science and the Arts and coordinated by the Bavarian Research Institute for Digital Transformation (bidt).
% \end{acks}

% \begin{acks}
%     We acknowledge that we used AI to...
% \end{acks}

%% Bibliography
\bibliographystyle{ACM-Reference-Format}
\bibliography{references}

\appendix
\section{Appendix}

\subsection{Appendix A: Interview Guide}

\subsection*{Pre-interview}
\begin{enumerate}
  \item Participant joins the study.
  \item Writes essay.
  \item The participant is asked to join a link to the interview Zoom meeting.
\end{enumerate}

\subsection*{Interview}

\paragraph{Introduce}
Hello, I am [Interviewer Name]. Thank you for joining our study and writing the essay!

\paragraph{Explain cued retrospective protocols}
In this task, we will go through your process of writing the piece you just wrote. Here you will have to try and recall what you were thinking while writing/doing what we see in the process replay. I only want you to tell me what you were doing and what you were thinking. You don’t need to justify any of the decisions you made here; you only have to recall and describe what you were thinking and doing at that moment.

During this, we will stop the replay at multiple points and try to remember what you were thinking at that particular moment in the writing process, what contributed to you writing the text you had written and so on. The goal of this exercise is to understand your process of writing and the decisions you made while writing.

\paragraph{Sample questions for eliciting retrospective protocols cued using the replay tool}
\begin{itemize}
  \item Do you remember what you were thinking at this point?
  \item Do you remember what you were thinking when this suggestion appeared?
\end{itemize}

\paragraph{Sample questions at the start of the session}
\begin{itemize}
  \item Did you have any concrete plans or ideas about what you wanted to write in this essay?
  \item How different/similar was the initial suggestion to your plan?
  \item Did you consider this suggestion and compare it with your ideas?
\end{itemize}

\paragraph{Sample questions when a suggestion appears}
\begin{itemize}
  \item What were you (already) thinking when the suggestion appeared? (Original thought train when suggestion appeared)
  \item What were you planning to write here? (Original plan before suggestion appeared)
  \item How much of the suggestion did you read?
  \item What did the suggestion make you think? (Thoughts that the suggestion triggered)
  \item What do you think of the suggestion? (Thoughts about the suggestion content itself)
\end{itemize}

\paragraph{Sample questions when writers have accepted a suggestion}
\begin{itemize}
  \item Why did you accept this suggestion?
  \item What were you planning to write here instead?
  \item How was the suggestion different from what you were planning to write?
  \item Why did you not select the remaining suggestion?
\end{itemize}

\paragraph{Sample questions when writers have written things after the suggestion}
\begin{itemize}
  \item Where did you get the idea for this sentence? (previous sentence, suggestion?)
  \item Can you compare the previous suggestion with this sentence you have written? (let them point out possible similarities and ask if there was any inspiration?)
\end{itemize}

\paragraph{End of the session}
\begin{itemize}
  \item How do you think the system generated the suggestions? (theory on the functioning of the suggestion system)
  \item Can you give me an example of what made you think this?
\end{itemize}

\paragraph{Possible themes to explore}
\begin{itemize}
  \item Inspiration for themes
  \item Intensity of expression
  \item Reasons for selection
\end{itemize}

\paragraph{Ethics and Consent}
At any point of the interview if you feel like you don't want to answer a particular question or want to stop the session, you can let me know. All data recorded during this session will only be used for the purpose of the study and will be deleted in 1 year after this session. All the data will also be anonymised and your identity will not be revealed.

\subsection{Appendix B: Dendrogram and 20 Final Topics}

\begin{figure*}[htbp]
    \centering
    \includegraphics[width=0.95\textwidth]{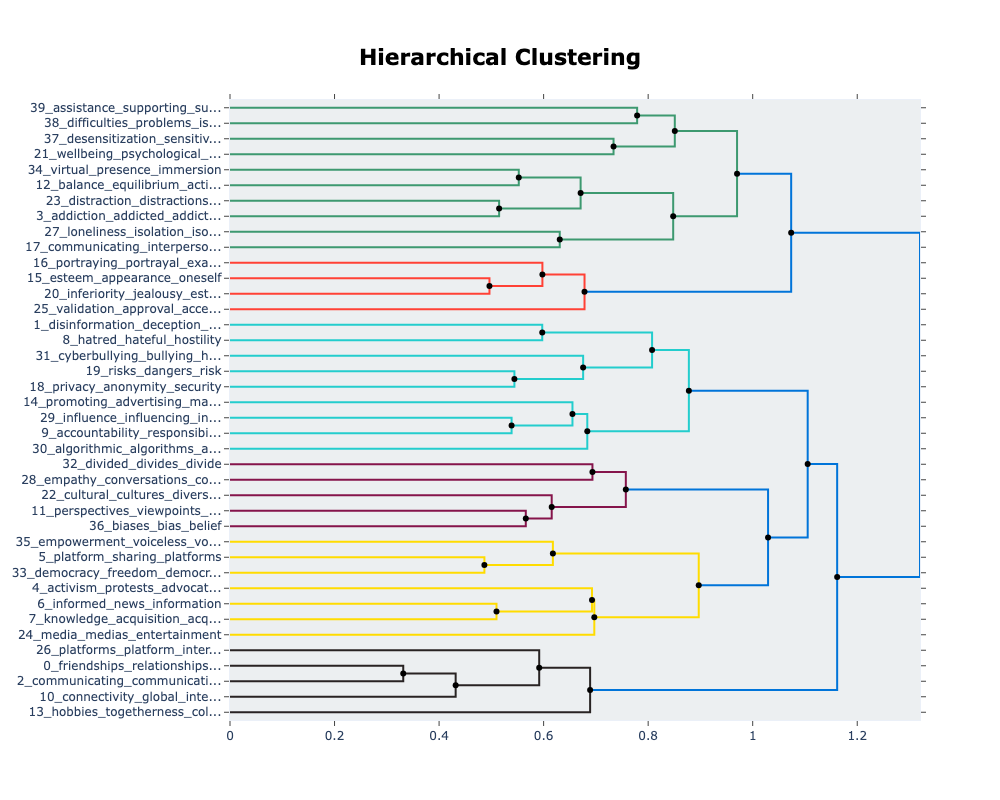}
    \caption{Dendrogram for merging and splitting topics.}
    \Description{Hierarchical clustering dendrogram showing the agglomerative merging of initial LLM-generated topics into the final 20 topics. The tree structure displays topic labels at the leaves with horizontal lines indicating merge points at different distance thresholds.}
    \label{fig:dendogram}
\end{figure*}

\begin{table}[t]
\centering
\small
\begin{tabular}{p{0.38\linewidth} p{0.55\linewidth}}
\hline
\textbf{Topic} & \textbf{Description} \\
\hline
addiction and distraction & Content about excessive use, need for moderation, inability to disconnect, or diverting attention from important activities \\
information sharing, news and learning & Content about acquiring knowledge, staying informed, or educational aspects \\
cyberbullying, arguments, and negativity & Content about online harassment, conflicts, or negative interactions \\
global connectivity, communication and networking (broader) & Content about connecting with new people, professional networking, or communication at a societal/global scale \\
interpersonal connections with friends and family (personal) & Content about MAINTAINING or STRENGTHENING existing personal relationships \\
mental health & Content about psychological well-being, emotional states, or mental conditions \\
misinformation, polarization, and trust & Content about false information, divided opinions, or issues of credibility \\
self-esteem, narcissism, and comparison to others & Content about self-image, excessive self-focus, or measuring oneself against others \\
activism, social movements, and awareness & Content about advocacy, collective action, or raising consciousness about issues \\
online anonymity & Content about identity concealment, pseudonymity, or operating without revealing one's identity \\
loneliness and isolation & Content about REDUCTION, LOSS, or ABSENCE of personal connections \\
financial opportunities and advancement & Content about economic benefits, career growth, or monetary aspects \\
online oversight and regulation & Content about rules, governance, or control of online spaces \\
privacy & Content about personal data protection, surveillance, or maintaining boundaries \\
discussion, debate, and diverse perspectives & Content about exchanging ideas, having conversations, or sharing different viewpoints \\
online business practices & Content about commercial activities, marketing, or corporate behavior online \\
empowering voices and free expression & Content about enabling speech, representation, or the ability to share one's thoughts \\
content on social media & Content specifically about the types of posts, videos, or material shared on platforms (not the impact of that content) \\
algorithms & Content about automated systems, recommendation engines, or computational processes \\
accountability and transparency & Content about responsibility, openness, or visibility of actions and decisions \\
uncategorized & \\
\hline
\end{tabular}
\caption{Topics and brief descriptions used for labeling (no description for uncategorized).}
\label{tab:topics}
\end{table}

\end{document}